\documentclass[sigconf]{acmart}
\usepackage{algorithm}
\usepackage{algorithmic}
\usepackage{amsmath} 
\usepackage{multirow}
\usepackage{float}
\AtBeginDocument{%
  }

\setcopyright{acmlicensed}
\copyrightyear{2018}
\acmYear{2018}
\acmDOI{XXXXXXX.XXXXXXX}
\acmConference[Conference acronym 'XX]{Make sure to enter the correct
  conference title from your rights confirmation email}{June 03--05,
  2018}{Woodstock, NY}
\acmISBN{978-1-4503-XXXX-X/2018/06}

\begin{document}

% \title{Content-based Enhancement for Personalized Product Search with Multimodal Large Language Models}

\title{Harnessing Multimodal Large Language Models for Personalized Product Search with Query-aware Refinement}

\author{Beibei Zhang}
\affiliation{%
  \institution{State Key Laboratory for Novel Software Technology, Nanjing University}
  \city{Nanjing}
  \country{China}
}
\email{zhangbb@smail.nju.edu.cn}

\author{Yanan Lu}
\affiliation{%
  \institution{Tencent}
  \city{Beijing}
  \country{China}
}
\email{yananlu@tencent.com}

\author{Ruobing Xie}
\affiliation{%
  \institution{Tencent}
  \city{Beijing}
  \country{China}
}
\email{xrbsnowing@163.com}

\author{Zongyi Li}
\affiliation{%
  \institution{Huazhong University of Science and Technology}
  \city{Wuhan}
  \country{China}
}
\email{zongyili@hust.edu.cn}

\author{Siyuan Xing}
\affiliation{%
  \institution{Tencent}
  \city{Beijing}
  \country{China}
}
\email{siyuanxing@tencent.com}

\author{Tongwei Ren}
\authornote{Corresponding author.}
\affiliation{%
  \institution{State Key Laboratory for Novel Software Technology, Nanjing University}
  \city{Nanjing}
  \country{China}
}
\email{rentw@nju.edu.cn}

\author{Fen Lin}
\affiliation{%
  \institution{Tencent}
  \city{Beijing}
  \country{China}
}
\email{felicialin@tencent.com}

\settopmatter{authorsperrow=4}
\renewcommand{\shortauthors}{Beibei Zhang et al.}

\begin{abstract}   
Personalized product search (PPS) aims to retrieve products relevant to the given query considering user preferences within their purchase histories. 
Since large language models (LLM) exhibit impressive potential in content understanding and reasoning, current methods explore to leverage LLM to comprehend the complicated relationships among user, query and product to improve the search performance of PPS.
Despite the progress, LLM-based PPS solutions merely take textual contents into consideration, neglecting multimodal contents which play a critical role for product search.
Motivated by this, we propose a novel framework, HMPPS, for \textbf{H}arnessing \textbf{M}ultimodal large language models (MLLM) to deal with \textbf{P}ersonalized \textbf{P}roduct \textbf{S}earch based on multimodal contents.
Nevertheless, the redundancy and noise in PPS input stand for a great challenge to apply MLLM for PPS, which not only misleads MLLM to generate inaccurate search results but also increases the computation expense of MLLM.
To deal with this problem, we additionally design two query-aware refinement modules for HMPPS:  
1) a perspective-guided summarization module that generates refined product descriptions around core perspectives relevant to search query, reducing noise and redundancy within textual contents;
and 2) a two-stage training paradigm that introduces search query for user history filtering based on multimodal representations, capturing precise user preferences and decreasing the inference cost.
Extensive experiments are conducted on four public datasets to demonstrate the effectiveness of HMPPS.
Furthermore, HMPPS is deployed on an online search system with billion-level daily active users and achieves an evident gain in A/B testing.

\end{abstract}


% %% the work being presented. Separate the keywords with commas.
% \keywords{Do, Not, Us, This, Code, Put, the, Correct, Terms, for,


\maketitle

\section{Introduction} 
  
\begin{figure}[!t]    
  \centering    
  \includegraphics[width=0.45\textwidth]{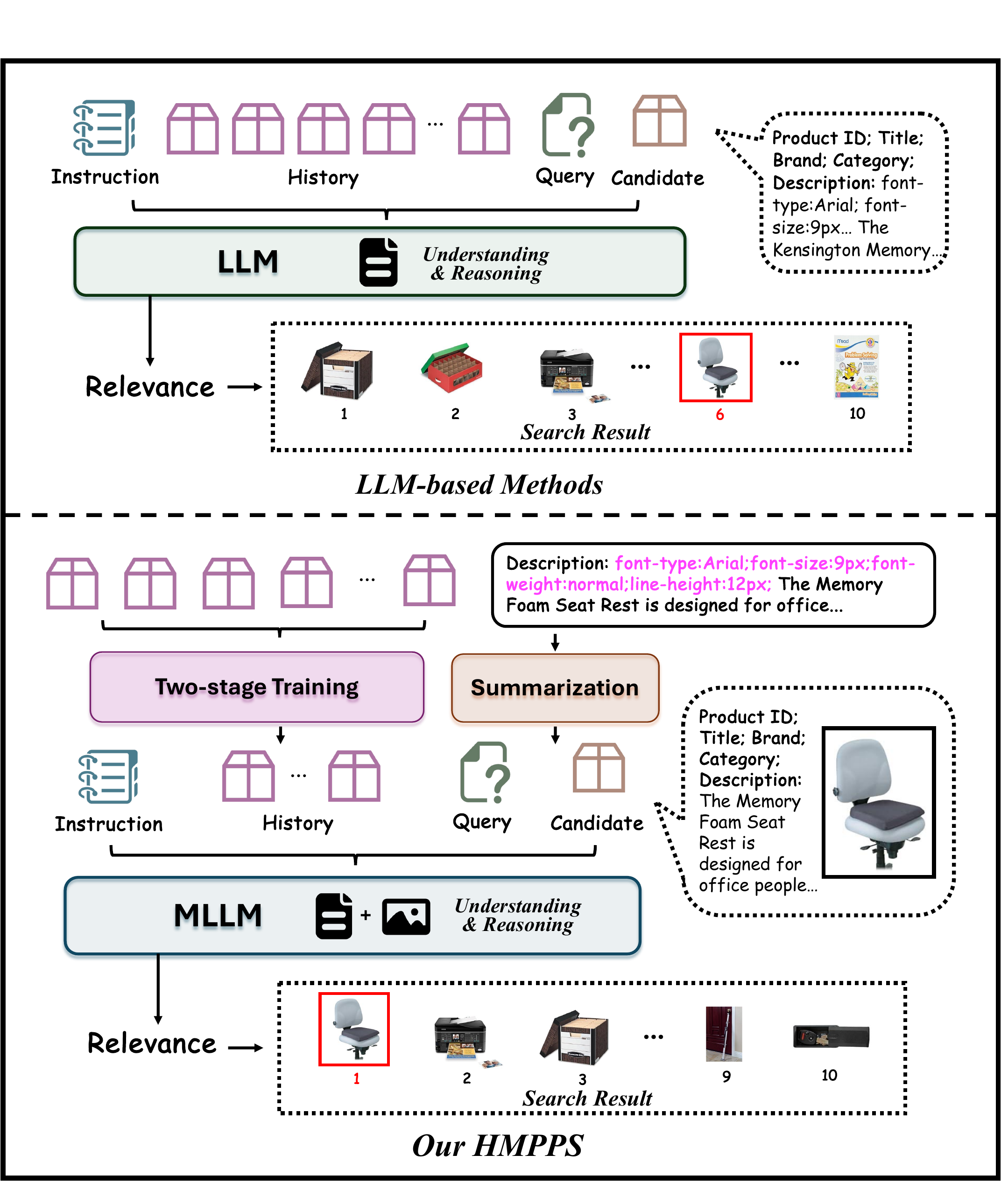}
  \caption{Comparison between existing LLM-based methods \emph{vs.} HMPPS. Here, the target product is labeled with a red box.}
\label{fig:motivation}  
\vspace{-0.45cm}   
\end{figure}  

Product search aims to return a ranked list of products in response to the given query submitted by users, which plays a vital role in online shopping services.
However, since limited queries are too ambiguous to effectively express the underlying preferences of users, the returned search results cannot satisfy user purchase intents precisely. To deal with this problem, more and more researches focus on personalized product search, which additionally takes user purchase history into consideration to model user preferences, contributing to capturing the exact purchase intents of users~\cite{ai2017learning,ai2019zero,ai2019explainable,ai2021model,liu2022category,shi2024unisar,bi2021learning, zhang2023recommendation,guo2023query}.

User, product and query are three core concepts in PPS. 
How to effectively represent the three and construct their complex relationships comprise the primary challenges of PPS. 
The primary practice in PPS is to convert user and product IDs into embedding vectors with trainable parameters and then conduct interaction modeling to predict the relevance among them~\cite{ai2017learning,xiao2019dynamic, jiang2020end, guo2023query, shi2024unisar}.
ID-based solutions are adept at learning the inherent pattern from the abundant log data for PPS and their limited-size embeddings contribute to the storage and computation efficiency.
Nevertheless, highly relying on ID-based data results in an inadequate understanding of the valuable content-based information of PPS, leading to a sub-optimal search result~\cite{ren2024representation}. 
Moreover, the negative impact of data bias, \emph{e.g.,} popularity bias~\cite{chen2023bias}, is inevitable for limited shopping logs. 
As a result, the low frequency and new unseen samples are necessarily insufficiently learned, harming PPS accuracy ultimately~\cite{yuan2023go, shen2024seminar}.

To make up for the weakness of ID-based methods, there have been some endeavors to exploit content-based information for PPS~\cite{bi2021learning, guo2018multi, wei2024towardsum}.
Particularly noteworthy is that the emergence of large language models, which exhibits revolutionized contribution to natural language processing, has inspired researchers to leverage their remarkable content understanding and reasoning capabilities to get rid of data restriction for PPS~\cite{zhang2023recommendation}.
As shown in the top half of Figure ~\ref{fig:motivation}, these approaches convert PPS into a language understanding task to predict the relevance among user history, query and product, which serves as a reranking accordance for a group of limited candidate products.
Despite the progress, LLM-based PPS methods achieve sub-optimal search performance for PPS since they merely depend on textual contents, neglecting multimodal information which play a critical role for product search. Visual display figures can be regarded as a powerful complement for PPS especially in cases of inadequate textual descriptions.

Motivated by these observations, we propose a novel framework, HMPPS, for harnessing multimodal large language models to deal with PPS based on multimodal contents.
As the bottom part of Figure ~\ref{fig:motivation} shows, we convert PPS into a multimodal language understanding task, which utilizes various multimodal contents to predict the relevance among user history, query and candidate product.
The remarkable multimodal content understanding ability of pre-trained MLLM contributes to processing and reasoning among diverse multimodal contents, achieving a thorough comprehension for PPS.
Leveraging HMPPS to rerank the search results filtered by existing ID-based methods, the entire search performance of PPS can be effectively enhanced.

Nevertheless, the redundancy and noise in PPS input stand for a great challenge to apply MLLM for PPS, which not only misleads MLLM to generate inaccurate search results but also increases the computation expense of MLLM.
Useless information in product descriptions, \emph{e.g.,} font settings in Figure ~\ref{fig:motivation}, are unhelpful for MLLM to understand the relations between product and query, which may even result in mistakes and hallucination.
Similarly, irrelevant purchased products in user history can cause mistakes in
comprehending user preferences for current search. 
Except for the accuracy reduction, limited context size and high computation cost of MLLM make it a huge burden to process overlong sequences caused by redundancy.

To address these issues, we additionally design two query-aware refinement modules for product description and user history refinement: 
1) a perspective-guided description summarization module that leverages an efficient LLM to obtain core perspectives relevant to search query and then summarize product descriptions around these perspectives, reflecting user search preferences;
and 2) a two-stage training paradigm that trains HMPPS for two stages where the first stage is trained on random user history to implicitly learn the correlation among user history, query and candidate product. The second stage is trained on selected user history filtered by the relevance between query and product multimodal representations extracted by the first-stage model, capturing more precise user preferences.
Both of these two modules contribute to not only improving the input robustness but also relieving the inference cost of HMPPS since the input size is decreased with the refined description and limited user history.

We conduct extensive experiments on four public datasets to demonstrate the effectiveness of HMPPS. 
The experimental results confirm the prominent advantage of HMPPS in enhancing the search performance of PPS. 
We also deploy HMPPS on an online search system with billion-level daily active users and achieve an evident gain in A/B testing, which validates the practicability of HMPPS.
It is worth noting that, profiting from the remarkable generalization of MLLM, training HMPPS using small-scale pre-trained MLLMs (\emph{e.g.,} InternVL2-1B\cite{chen2024internvl}) on small-scale training samples (\emph{e.g.,} 10\% of the entire training set) can achieve an obvious improvement, proving the training efficiency of HMPPS.

The main contributions of our work can be summarized as follows:

\begin{itemize}
  \item We propose a novel method, HMPPS, which utilizes pre-trained MLLMs to deal with PPS based on multimodal contents, enhancing the entire search performance of PPS.
  \item We design a perspective-guided description summarization module for HMPPS, utilizing LLM to generate refined summaries around core perspectives relevant to search query, reducing the redundancy and noise in data.
  \item We design a two-stage training paradigm for HMPPS to obtain limited user history relevant with the input query and candidate product, which improves the search accuracy and decreases the inference cost of HMPPS.
\end{itemize}

\begin{figure*}[!t] 
  \centering 
  \includegraphics[width=1\textwidth]{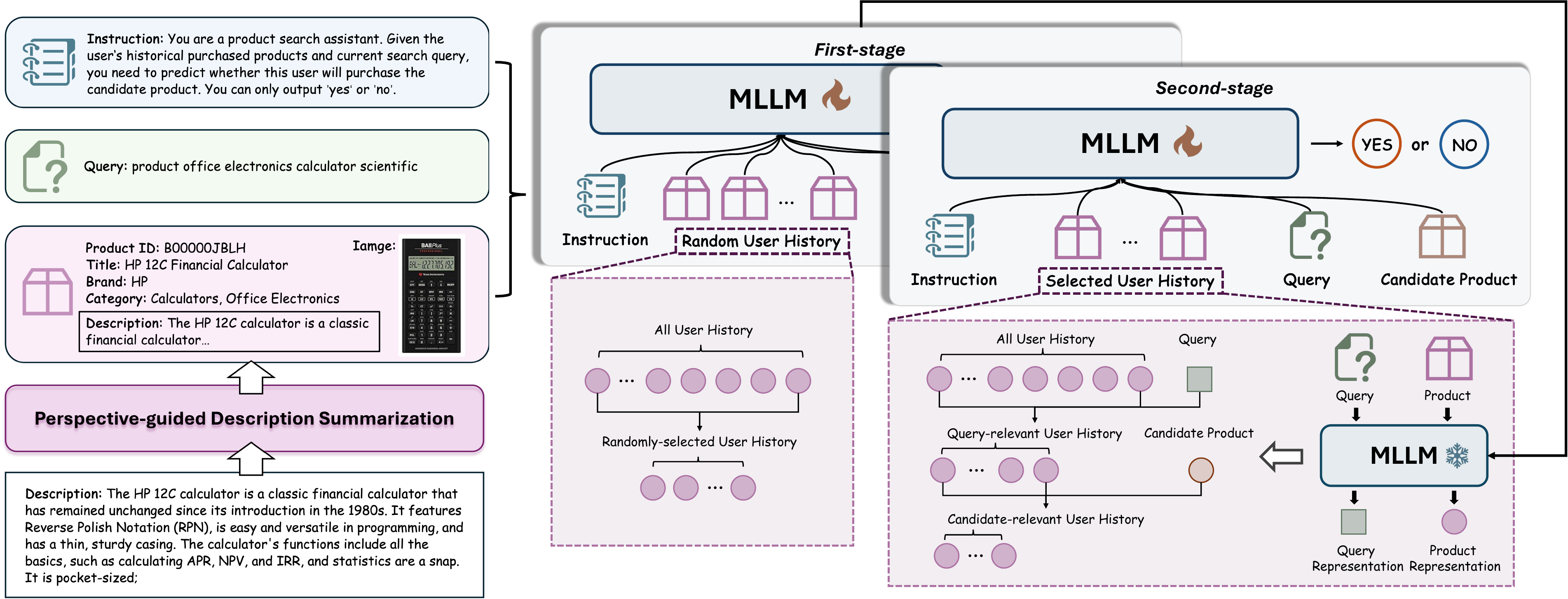}
  \caption{An overview of HMPPS. User history, query and candidate product are fed into MLLM to generate the search decision guided by the designed instruction based on abundant multimodal contents. The product description is additionally refined with a perspective-guided description summarization module. During training, for the first stage, the MLLM is trained to learn the relationship between query and user history implicitly, which is then leveraged to extract product and query representations for user history selection. For the second stage, this MLLM is further trained with selected query-relevant user history for more accurate prediction.}
\label{fig:framework}
\end{figure*}

\section{Related Work}
\textbf{Personalized Product Search.}
Previous PPS solutions tend to convert user, product and query IDs to embedding vectors and then predict the relevance among them using various interaction modeling methods~\cite{ai2017learning, ai2019explainable, xiao2019dynamic, ai2019zero, jiang2020end, liu2020structural, bi2020transformer, bi2021learning, liu2022category,cheng2022ihgnn, thonet2022joint, fan2022modeling, wu2023dynamic}. 
Since ID-based methods heavily rely on dataset quality, inevitable data bias caused by the limited dataset collection results in insufficient learning of low frequency samples.
As a result, several methods utilize content-based information to deal with these problems from semantic aspects~\cite{bi2021learning, guo2018multi, wei2024towardsum}, where LLM-based solutions exhibit significant superiority due to the remarkable content understanding and reasoning capabilities within LLMs~\cite{zhang2023recommendation}.
Nevertheless, LLM-based methods merely depend on textual contents, neglecting multimodal information which play a critical role for PPS.
The proposed HMPPS is targeted for addressing this issue by introducing MLLM to deal with PPS based on multimodal contents.

Some PPS methods~\cite{ai2019zero, guo2019attentive, pan2021personalized,wu2022meta, shen2022hierarchically, dai2023contrastive, shi2024unisar} commit to conduct complicated user history modeling to extract precise user preferences, taking lifelong historical behaviors as model input, which is impractical for MLLM application with limited context size. 
Most relevant to our work is QIN~\cite{guo2023query}, which designs a cascaded strategy to filter irrelevant historical products via product and query representation matching. 
However, off-the-shelf representation extractors in QIN cannot well adapt to PPS domain, weak in understanding the relations among product and query in search scenario.  
Our two-stage training paradigm not only extracts product and query representations using a powerful finetuned MLLM but also captures more precise relations between query and product based on multimodal information.

\noindent\textbf{Large Language Models for Description Summarization.}
Large language models have emerged as powerful tools in the field of natural language processing, for which more and more methods propose to leverage LLMs for product description summarization to reduce noise and redundancy~\cite{torbati2023recommendations, ren2024representation, wu2025tfdcon}.
To prevent hallucinations in generated summaries, some methods~\cite{xi2024towards}, motivated by information factorization, leverage manually collected factors to guide LLM summarization, which results in high labor cost and limited generalization ability.

Different from these methods, the perspective-guided description summarization module in HMPPS leverages LLM to automatically extract core perspectives based on product descriptions and search queries, reflecting user search preferences.
Additionally, we apply chain of thought and one-shot demonstration to ensure the reliability of the extracted perspectives and generated summaries.

\section{Problem Formulation}
\vspace{0.1cm}
PPS aims to retrieve products relevant to the given query considering user preferences within their purchase histories.
Let ${U}$, ${P}$ and ${Q}$ denote the sets of users, products and queries, respectively. Each user $u$ has a chronologically ordered purchase history which is composed of products $H_u = \{ p^u_1, p^u_2, ..., p^u_{N_u} \}$, where $N_u$ stands for the number of previously purchased products.
The target of PPS is to predict the probability of $u$ purchasing the candidate product $p_i \in P$ in response to the given query $q$ based on his/her purchase history $H_u$:
{\begin{equation}
    {y_{u,q,p_i}} = \mathcal{F}_\theta \left(H_u, q, p_i\right),
\label{con:problem_formulation}
\end{equation}
}  
where $y_{u,q,p_i}$ indicates the purchase probability, and $\mathcal{F}$ refers to the PPS solution with learnable parameters $\theta$.

The standard PPS output is a ranking list of all products in ${P}$ according to their purchased probabilities obtained by Equation ~\ref{con:problem_formulation}.
However, HMPPS aims to capture exact but costly fine-grained relationships while ID-based methods can efficiently generate massive relevance scores with limited-size user, product and query embeddings.
As a result, we decide to obtain top-$K_p$ candidates ${P}^\prime = \{ p^\prime_1, p^\prime_2, ..., p^\prime_{K_p}\}$, where $K_p \ll N_P$, $N_P$ denotes the product number of ${P}$, based on the search results of an existing ID-based method $M_{ID}$.
And HMPPS plays the role of reranker that predicts $y_{u,q,p_i}$ only for $p_i \in {P}^\prime$, which is a common practice in content-based PPS~\cite{bi2021learning, zhang2023recommendation}.
This formulation combines the complementary advantages of HMPPS and ID-based methods, boosting the entire PPS performance.

\section{Method}
\vspace{0.2cm}
The overall framework of HMPPS is shown in Figure ~\ref{fig:framework}, which consists of three main components: 1) MLLM-based PPS. With instructed prompts, we convert PPS into a multimodal language understanding task and leverage MLLM to predict the relevance among user history, query and product based on multimodal contents, which comprises the search backbone of HMPPS; 2) perspective-guided description summarization. We firstly collect search-relevant perspectives based on product descriptions and search queries and then conduct summarization according to these limited perspectives with the help of LLM, which serves as one input refinement module for HMPPS;
and 3) two-stage training paradigm. We firstly train MLLM with random user history to implicitly learn the correlation among user, query and product. 
The first-stage finetuned MLLM is then leveraged to select historical products relevant to search query and candidate product, serving for the second-stage training to earn a more accurate prediction. This paradigm can be regarded as the other input refinement module for HMPPS.

\vspace{0.1cm}
\subsection{MLLM-based PPS}
\label{sec:mllm_backbone}
\vspace{0.2cm}
As shown in Figure ~\ref{fig:framework}, we design a specific template to transform the point-wise reranker of PPS into MLLM formulation by aggregating instruction text $Inst$, user history $H_u$, query $q$ and candidate product $p_i$ into an instructed prompt, which enforces the purchase decision output $d$ to be either ``yes'' or ``no''. 
Since HMPPS concentrates on adequately mining valuable contents for PPS enhancement, we take various multimodal information to represent product, including textual product ID, title, brand, category, description and visual display figure. 
We leverage the general language generation loss to finetune the pre-trained MLLM model, migrating its powerful understanding and reasoning ability to the PPS domain:

{\begin{equation} 
    {\mathcal{L}} = - \sum_{l=1}^{L} \text{logPr}(d_l|d_{<l},[Inst;H_u;q;p_i]),
  \label{con:training_Loss}
  \end{equation}
  } 
where $d_l$ is the $l$-th word in the decision text, $L$ is the length of the decision text, and $\text{Pr}(\cdot)$ denotes the next word probability predicted by HMPPS.

Contrast learning based on positive and negative samples is proved to be an effective solution to boost the ranking accuracy~\cite{ma2024finetuning}. 
As a result, for each search log $\langle H_u, q \rangle$, except for the target product, we randomly sample $K^n_s$ candidate products from the entire product set ${P}$ as simple negatives.
Moreover, we obtain $K^n_h$ hard negatives from the search result ${P}^\prime$ of existing ID-based method $M_{ID}$ according to their predicted relevance scores.
All positive and negative candidates participate in the HMPPS optimization of Equation \ref{con:training_Loss} by assigning ``yes'' as the positive sample decision and ``no'' as the negative sample decision.

During the inference stage, since the target of PPS is to predict the relevance score instead of a discrete token, we intercept the probability distribution of the predicted next word and conduct a bidimensional softmax over the estimated scores corresponding to ``yes'' and ``no'' to obtain the final purchase probability $y_{u,q,p_i}$:

{\begin{equation}
      {y_{u,q,p_i}} = \frac{\text{exp}(s_{yes})}{\text{exp}(s_{yes}) + \text{exp}(s_{no})},
  \label{con:infer}
  \end{equation}
  }
where $s_{yes}$ and $s_{no}$ refer to the estimated scores related to ``yes'' and ``no'' in the predicted probability distribution of the output word, respectively.

\subsection{Perspective-guided Description Summarization}
 
Factorization has been proved to be an efficient approach to reduce hallucination for information refinement~\cite{xi2024towards}, motivated by which we design a perspective-guided summarization module that utilizes LLM to summarize product descriptions with two relative prompts: 
1) perspective extraction prompt that finds out perspectives that users concern for product search by exploring the relationships between the product and search query;
and 2) summary generation prompt that instructs LLM to summarize product descriptions concentrating on the search-relevant perspectives collected by the previous step.
Except for hallucination reduction, by introducing search-relevant perspectives, the generated summaries can reflect user search preferences more accurately, pandering to the PPS target.

\subsubsection{Perspective Extraction Prompt.}
In-context learning is an effective paradigm to improve LLM generation performance with a few demonstrations composed by example inputs and outputs~\cite{brinkmann2023product}. 
Therefore, except for the task instruction, we append one demonstration to the perspective extraction prompt to constraint the LLM output and boost the generation quality.

After collecting perspectives from all samples, we retain the top-$K_d$ frequent perspectives as core perspectives that users concern most during product search.
Since user search preferences vary in different scenarios, we provide demonstration and collect perspectives respectively for each dataset.

\subsubsection{Summary Generation Prompt.}
We employ LLM to summarize product descriptions centering on the collected core perspectives to obtain refined descriptions which are more correlative with user search preferences.
Since description summarization is a challenging language processing task, we utilize chain of thought to reduce LLM hallucination~\cite{wei2022chain}, enhancing robustness of the generated summary. 
To be specific, except for summary generation, we explicitly include a reasoning demand in task instruction to enforce LLM to execute the reasoning procedure before outputting summarization result.
One-shot demonstration is also available here to help LLM better understand the complicated summarization task.

The summarized descriptions are regarded as the product description of MLLM input to take part in the PPS prediction in section ~\ref{sec:mllm_backbone}.
Even though the summarization module involves LLM inference for two steps, it can be conducted off-line for only once and the summarized results can be saved for future use, which yeilds immeasurable benefits with negligible expense.

\subsection{Two-stage Training Paradigm}
User preferences tend to vary for different query and candidate product~\cite{ai2019zero}. 
To capture exact user preferences, it is necessary to stand out correlative purchased products from the entire user history exclusively with the given query and candidate product.
However, due to the limited context size and high computation cost, it is impractical to feed the entire set of historical products, together with query and candidate product, into MLLM to directly capture prominent purchased products.
Consequently, we propose a two-stage training paradigm to efficiently extract precise user history.

In the first stage, to cover the lifelong sequential user history with MLLM of limited input size, we randomly select $K_{s_1}$ chronological purchased products from the entire user history $H_u$, which subsequently take part in the MLLM training with the PPS optimization target in Section ~\ref{sec:mllm_backbone}.
The transformer architecture of MLLM is adept at capturing dynamic fine-grained relationships among user history, query and candidate product, assigning higher attention to relevant historical products and lower attention to irrelevant historical products.
Thus, after the first-stage optimization, the finetuned MLLM has learned the implicit relevance between products and queries for the specific search scenario.
We then apply this MLLM as encoder to extract multimodal representations of each product and query by averaging embeddings of all output tokens:
{
  \begin{align}
    & f_p = \text{Avg} \left( \mathcal{F}_{MLLM} \left( [ t_p; v_p ]; \theta^{s_1} \right) \right), \\
    & f_q =\text{Avg}\left(\mathcal{F}_{MLLM}\left(t_q; \theta ^ {s_1}\right)\right), 
  \end{align}
}
where $\text{Avg}$ denotes the average operation on embeddings, $\mathcal{F}_{MLLM}$ denotes the MLLM architecture with the first-stage learned parameters $\theta^{s_1}$, $t_p$ is the combination of product textual information, including product title, brand, category and description, $v_p$ is the product display figure and $t_q$ refers to the query.
Cosine similarity is calculated based on the extracted representations to obtain the product-query and product-product relevance:

{
  \vspace{-0.1cm}
  \begin{align}
    r_{p^u_jx} &= cos(f_{p^u_j}, f_x) \\
    & = \frac{f_{p^u_j} \cdot f_{x}}{\Vert f_{p^u_j} \Vert \Vert f_{x} \Vert}, x \in \{q, p_i\}, \nonumber
  \end{align}
} 
where $cos(\cdot)$ denotes the cosine similarity calculation between vectors and $\Vert \Vert$ denotes the euclidean norm of vectors.

According to the relevance $r_{p^u_jq}$ between historical product and query, we firstly collect $2K_{s_2}$ products from the entire historical product set $H_u$ to construct a query-relevant user history set $H^\prime_u$. 
Then we pick out $K_{s_2}$ products from $H^\prime_u$ according to $r_{p^u_jp_i}$ as the final selected user history set $H_u^{\prime\prime}$, which takes the relevance between historical product and candidate product into consideration.
    
We take $H_u^{\prime\prime}$ as the user history input to further train the MLLM for the second stage of HMPPS, which learns more precise relationships among the relevant user history, query and candidate product.
On one hand, a more accurate ranking result can be generated due to the irrelevant historical products have been filtered out.
On the other hand, the product number of user history has been decreased for $K_{s_2} < K_{s_1} < N_u$, which reduces the final inference cost of HMPPS. 

During inference, since product and query representations have been extracted by the first-stage MLLM and saved in advance, user history selection can be directly conducted via calculating representation similarity.
The selected historical products are then regarded as user history to feed the second-stage MLLM to generate the final search decision.

\section{Experiments}
We conduct extensive experiments on multiple public available datasets to prove the effectiveness of HMPPS. Specifically, our experiments are intended to answer the following research questions (RQs):
\begin{itemize}
  \item \textbf{RQ1} Can HMPPS improve the entire search performance for PPS by reranking the candidate products filtered by existing ID-based methods?
  \item \textbf{RQ2} Does HMPPS outperform other content-based PPS solutions?
  \item \textbf{RQ3} Does each component of HMPPS take effect?
  \item \textbf{RQ4} Does HMPPS have the potential for boosting PPS performance with different training scales? 
  \item \textbf{RQ5} How does HMPPS make up for the limitations of existing PPS approaches?
  \item \textbf{RQ6} Can HMPPS benefit the real-world online search system? 

\end{itemize}

\begin{table*}[t!]\footnotesize
  \begin{center}
  \setlength{\tabcolsep}{1.6mm}
  \caption{Comparison results of PPS performance improvement by utilizing HMPPS to rerank the candidate products filtered by different ID-based PPS methods on all datasets. Here, M, N and R denote the metrics of MRR, NDCG and Recall, respectively, Rel.Impr refers to the relative improvement rate of HMPPS against the basic search approaches among all metrics, and the best result is in bold.} 
  \renewcommand{\arraystretch}{1.1}
    \begin{tabular}{c|cccc|cccc|cccc|cccc}
  \toprule
  \multirow{2}{*}{Model} & \multicolumn{4}{c|}{ Office Products } & \multicolumn{4}{c|}{ Cell Phones \& Accessories } & \multicolumn{4}{c|}{ Beauty } & \multicolumn{4}{c}{ Sports \& Outdoors } \\
  \cmidrule{2-17}
  ~ & M@8  & N@4 & R@4  & R@1 & M@8  & N@4 & R@4  & R@1 & M@8  & N@4 & R@4  & R@1 & M@8  & N@4 & R@4  & R@1 \\
  \midrule
  HEM & 0.243 & 0.239 & 0.344 & 0.114 
  & 0.081 & 0.078 & 0.109 & 0.040  
  & 0.137 & 0.132 & 0.180 & 0.073 
  & 0.169 & 0.167 & 0.227 & 0.096  \\
  + HMPPS  & \textbf{0.288} & \textbf{0.292} & \textbf{0.392} & \textbf{0.169} & \textbf{0.101} & \textbf{0.104} & \textbf{0.136} & \textbf{0.062} & 
  \textbf{0.155} & \textbf{0.157} & \textbf{0.208} & \textbf{0.091} & 
  \textbf{0.185} & \textbf{0.192} & \textbf{0.257} & \textbf{0.111} \\
  Rel.Impr & 18.52\%& 22.18\%& 13.95\%& 48.25\%
  & 24.69\%& 33.33\%& 24.77\%& 55.00\%
  & 13.14\%& 18.94\%& 15.56\%& 24.66\%
  & 9.47\%& 14.97\%& 13.22\%& 15.62\%\\
  \midrule
  ZAM & 0.222 & 0.216 & 0.303 & 0.111 
  & 0.069 & 0.065 & 0.092 & 0.033 
  & 0.101 & 0.098 & 0.135 & 0.052 
  & 0.103 & 0.101 & 0.139 & 0.056 \\
  + HMPPS  & \textbf{0.274} & \textbf{0.280} & \textbf{0.372} & \textbf{0.163} & \textbf{0.097} & \textbf{0.100} & \textbf{0.128} & \textbf{0.063} & 
  \textbf{0.134} & \textbf{0.137} & \textbf{0.175} & \textbf{0.086} & 
  \textbf{0.137} & \textbf{0.144} & \textbf{0.188} & \textbf{0.089} \\
  Rel.Impr & 23.42\%& 29.63\%& 22.77\%& 46.85\%
  & 40.58\%& 53.85\%& 39.13\%& 90.91\%
  & 32.67\%& 39.80\%& 29.63\%& 65.38\%
  & 33.01\%& 42.57\%& 35.25\%& 58.93\%\\
  \midrule
  DREM & 0.193 & 0.186 & 0.265 & 0.093 
  & 0.095 & 0.091 & 0.125 & 0.050 
  & 0.135 & 0.131 & 0.182 & 0.070 
  & 0.163 & 0.161 & 0.218 & 0.092 \\
  + HMPPS  & \textbf{0.240} & \textbf{0.246} & \textbf{0.330} & \textbf{0.141} & \textbf{0.106} & \textbf{0.108} & \textbf{0.145} & \textbf{0.061} & 
  \textbf{0.157} & \textbf{0.160} & \textbf{0.213} & \textbf{0.093} & 
  \textbf{0.182} & \textbf{0.189} & \textbf{0.253} & \textbf{0.111} \\
  Rel.Impr & 24.35\%& 32.26\%& 24.53\%& 51.61\%
  & 11.58\%& 18.68\%& 16.00\%& 22.00\%
  & 16.30\%& 22.14\%& 17.03\%& 32.86\%
  & 11.66\%& 17.39\%& 16.06\%& 20.65\% \\
  \midrule
  DREM-HGN & 0.241 & 0.232 & 0.314 & 0.130 
  & 0.066 & 0.062 & 0.086 & 0.034
  & 0.141 & 0.136 & 0.183 & 0.077 
  & 0.120 & 0.117 & 0.164 & 0.062  \\
  + HMPPS  & \textbf{0.262} & \textbf{0.269} & \textbf{0.365} & \textbf{0.147} & \textbf{0.080} & \textbf{0.081} & \textbf{0.106} & \textbf{0.049} & 
  \textbf{0.161} & \textbf{0.163} & \textbf{0.215} & \textbf{0.099} & 
  \textbf{0.154} & \textbf{0.160} & \textbf{0.211} & \textbf{0.098} \\
  Rel.Impr & 8.71\%& 15.95\%& 16.24\%& 13.08\%& 
  21.21\%& 30.65\%& 23.26\%& 44.12\%& 
  14.18\%& 19.85\%& 17.49\%& 28.57\%& 
  28.33\%& 36.75\%& 28.66\%& 58.06\% \\
  \midrule
  CAMI & 0.244 & 0.239 & 0.330 & 0.129 
  & 0.088 & 0.086 & 0.121 & 0.045 
  & 0.130 & 0.125 & 0.172 & 0.068 
  & 0.170 & 0.168 & 0.227 & 0.096 \\
  + HMPPS & \textbf{0.268} & \textbf{0.271} & \textbf{0.366} & \textbf{0.154} & \textbf{0.103} & \textbf{0.104} & \textbf{0.136} & \textbf{0.063} & 
  \textbf{0.156} & \textbf{0.159} & \textbf{0.210} & \textbf{0.095} & 
  \textbf{0.189} & \textbf{0.196} & \textbf{0.264} & \textbf{0.113} \\
  Rel.Impr & 9.84\%& 13.39\%& 10.91\%& 19.38\%& 
  17.05\%& 20.93\%& 12.40\%& 40.00\%& 
  20.00\%& 27.20\%& 22.09\%& 39.71\%& 
  11.18\%& 16.67\%& 16.30\%& 17.71\%\\
  \midrule
  UniSAR & 0.368 & 0.370 & 0.475 & 0.232 
  & 0.155 & 0.152 & 0.205 & 0.086
  & 0.130 & 0.125 &  0.173 & 0.064
  & 0.181 & 0.173 & 0.211 & 0.128\\
  + HMPPS & \textbf{0.398} & \textbf{0.414} & \textbf{0.532} & \textbf{0.256} & \textbf{0.190} & \textbf{0.194} & \textbf{0.249} & \textbf{0.124} & 
  \textbf{0.184} & \textbf{0.188} & \textbf{0.238 } & \textbf{0.121} & 
  \textbf{0.198} & \textbf{0.207} & \textbf{0.269} & \textbf{0.130} \\
  Rel.Impr & 8.15\%& 11.89\%& 12.00\%& 10.34\%& 
  22.58\%& 27.63\%& 21.46\%& 44.19\%& 
  41.54\%& 50.40\%& 37.57 \%& 89.06\%& 
  9.39\%& 19.65\%& 27.49\%& 1.56\% \\
  \bottomrule
  \end{tabular}
  \label{tab:performance_comparison}  
  \end{center}  
  \end{table*}

  \begin{table*}[t!]\footnotesize
    \begin{center}
    \setlength{\tabcolsep}{2.2mm}
    \caption{Comparison results of PPS performance of HMPPS \emph{vs.} other content-based PPS methods on all datasets.} 
    \renewcommand{\arraystretch}{1.1}
    \begin{tabular}{c|cccc|cccc|cccc|cccc}
    \toprule
    \multirow{2}{*}{Model} & \multicolumn{4}{c|}{ Office Products } & \multicolumn{4}{c|}{ Cell Phones \& Accessories } & \multicolumn{4}{c|}{ Beauty } & \multicolumn{4}{c}{ Sports \& Outdoors } \\
    \cmidrule{2-17}
    ~ & M@8  & N@4 & R@4  & R@1 & M@8  & N@4 & R@4  & R@1 & M@8  & N@4 & R@4  & R@1 & M@8  & N@4 & R@4  & R@1 \\
    \midrule
    RTM & 0.371 & 0.382 & 0.499 & 0.228 
    & 0.130 & 0.130 & 0.189 & 0.058
    & 0.143 & 0.146 & 0.204 & 0.074
    & 0.162 & 0.169 & 0.239  & 0.088 \\
    InstructRec  & 0.382 & 0.398 & 0.521 & 0.239
    & 0.184 & 0.187 & 0.241 & 0.117
    & 0.175 & 0.180 & 0.232 &  0.110
    & 0.171 & 0.178 & 0.244 & 0.098 \\
    HMPPS & \textbf{0.398} & \textbf{0.414} & \textbf{0.532} & \textbf{0.256} & \textbf{0.190} & \textbf{0.194} & \textbf{0.249} & \textbf{0.124} & 
    \textbf{0.184} & \textbf{0.188} & \textbf{0.238 } & \textbf{0.121} & 
    \textbf{0.198} & \textbf{0.207} & \textbf{0.269} & \textbf{0.130} \\
    \bottomrule
  \end{tabular}
  \label{tab:content_comparison}  
  \end{center}  
  \end{table*}

\vspace{-0.3cm}
\subsection{Experimental Setup}

\subsubsection{Datasets and Evaluation Metrics.}
We take 5-core Amazon product search dataset~\cite{mcauley2015inferring} as our experimental corpus. 
Following previous PPS works\cite{ai2017learning, ai2019zero, ai2019explainable}, we select four diverse subsets of the Amazon dataset to conduct experiments, which are \emph{Office Products} (Office), \emph{Cell Phones \& Accessories} (Cell), \emph{Beauty} and \emph{Sports \& Outdoors} (Sports). 
To evaluate the PPS performance of HMPPS, we utilize three typical search metrics which are Mean Reciprocal Rank (MRR), Normalized Discounted Cumulative Gain (NDCG) and Recall.

\vspace{-0.2cm}
\subsubsection{Implementation Details.}
We take InternVL2-1B~\cite{chen2024internvl} as the MLLM backbone.
The size of reranking candidate product set $K_p$ is set to 10. 
We utilize Qwen2.5-14B~\cite{qwen2.5} for description summarization and the number of core perspectives $K_d$ is set as 20.
The ID-based method $M_{ID}$ is UniSAR, generating the basic search results for HMPPS reranking.
The number of simple negative samples $K_s^n$ is 2 and the number of hard negatives $K_h^n$ is 3. 
The product number $K_{s_1}$ of randomly selected user history is set as \{5, 7, 7, 7\} for Office, Cell, Beauty and Sports datasets, respectively.
The product number $K_{s_2}$ of user history selected based on the relevance is set as \{2, 3, 3, 3\} for the second-stage of HMPPS.
Low-rank adaption (LoRA) is adopted for parameter-efficient finetuning MLLM. Benefitting from the remarkable generalization of MLLM, we train HMPPS on merely 10\% training data.

\vspace{-0.2cm}
\subsection{Performance Comparison (RQ1 \& RQ2)}
\label{sec:performance_comparison}
To prove the effectiveness of HMPPS for PPS, we integrate it with six different representative PPS solutions to rerank their search results.
The evaluation results are shown in Table ~\ref{tab:performance_comparison}, which reveals the following observations:
1) HMPPS can effectively utilize MLLM to rerank the candidate products filtered by existing PPS models based on multimodal contents, resulting in obvious improvement and enhancing the entire search performance for PPS;
and 2) HMPPS can adapt to various types of existing PPS models. The obvious enhancement among all metrics for all existing models verifies the generalization of HMPPS.

To verify the superiority of HMPPS in PPS content understanding, we compare it with other content-based solutions to rerank the candidate products filtered by UniSAR. 
As Table ~\ref{tab:content_comparison} shows, HMPPS performs better in content-based reranking compared to conventional and LLM-based solutions, proving the advantage of HMPPS in content understanding for PPS.
It is worth noting that InstructRec is trained on the full dataset based on a 3B language model while HMPPS is implemented based on a 1B model and trained for merely 10\% training data. The positive experimental results demonstrate that, except for the accuracy improvement, HMPPS can effectively alleviate the severe dependency on data and lead to an efficient training procedure.

In addition, as shown in Table ~\ref{tab:qin_comparison}, HMPPS outperforms QIN even though the basic retriever exhibits inferiority, which validates the effectiveness of HMPPS in capturing valid user history with the two-stage training paradigm.

\begin{table}[t]\small
  \begin{center}  
  \caption{Comparison results of PPS performance of HMPPS \emph{vs.} user history selection PPS method, QIN, on Beauty dataset. Here, HMPPS$^*$ stands for HMPPS trained on the full dataset. } 
  \renewcommand{\arraystretch}{1.1}
  \begin{tabular}{c|ccc}
    \toprule
     Model & MRR@8 & NDCG@4 & Recall@4 \\
    \midrule
     QIN  & 0.222 & 0.231 & 0.321  \\
     HMPPS & 0.234 &0.238 &0.315 \\
     HMPPS$^{*}$  & \textbf{0.257} &\textbf{0.263} &\textbf{0.340} \\
    \bottomrule
  \end{tabular}  
  \label{tab:qin_comparison}  
  \end{center} 
  \vspace{-0.1cm}
  \end{table} 

  \begin{table}[t]\small
    \begin{center}  
    \caption{Ablation results of HMPPS applying contents of different modalities as input on Office Products dataset.} 
    \renewcommand{\arraystretch}{1.1}
    \begin{tabular}{c|c|cccc}
      \toprule
       Content Type & Training  & M@8 & N@4 & R@4 & R@1 \\
      \midrule
       Text & Zero-shot & 0.203 & 0.189 & 0.287 & 0.074 \\
       Text + Vision & Zero-shot & 0.220 & 0.208 & 0.310 & 0.087   \\
       Text & Finetuned & 0.376 & 0.391 & 0.514 & 0.231  \\
       Text + Vision & Finetuned & 0.386 & 0.401 & 0.524 & 0.240  \\ 
      \bottomrule
    \end{tabular}  
    \label{tab:multimodal_ablation}  
    \end{center} 
    \vspace{-0.2cm}
    \end{table} 

\vspace{-0.3cm}
\subsection{Ablation Study (RQ3)}
\subsubsection{MLLM-based PPS}
To verify the advantage of utilizing MLLM for PPS based on multimodal contents, we compare the results of taking only textual information as input with those including visual figures. To reduce distraction, we keep the original textual descriptions without summarization and train HMPPS only for the first stage in this experiment.

As shown in Table ~\ref{tab:multimodal_ablation}, no matter whether the backbone model is finetuned or not, it is superior to apply the multimodal combination of textual and visual contents as HMPPS input, proving that multimodal contents contribute to search information complementary for PPS.
In addition, finetuned models consistently perform better than those without training, which validates that it is necessary to finetune MLLM on specific search corpus since PPS is a challenging task that requires knowledge migration.

\subsubsection{Perspective-guided Description Summarization.}
\label{description_summarization_ablation}
To explore the most appropriate summarization strategy, we design the other two types of prompts for comparison: 1) direct summarization prompt that instructs LLM to generate description summary directly without any demonstration and reasoning request; and 2) reasoning-based summarization prompt that requires LLM to reason before generating the final summary with one-shot demonstration, which can be seen as a simple version of perspective-guided summarization without an explicit declaration for specific perspectives. 
The comparison results among the original and all three types of summarized descriptions are shown in Table ~\ref{tab:description_summarization}.

The experimental results lead to three observations: 
1) utilizing LLM to refine product descriptions into decreased words will not negatively influence or even boost the PPS performance of HMPPS, which proves that product descriptions indeed contain redundancy and noise;
2) even though reasoning-based summarization prompt achieves the smallest average word count, its summarization result is not stable which results in the largest max word count and worst PPS performance.
Comparing to direct summarization prompt, the reasoning instruction increases the generation complexity for LLM while it lacks explicit perspectives that constrain the LLM output in perspective-guided summarization;
and 3) perspective-guided summarization prompt achieves the best PPS performance due to the reasonable summarization procedure around search-relevant perspectives with an acceptable processing cost, which becomes the final choice for description summarization.

\begin{table}[t]\small
  \begin{center}  
  \caption{Ablation results of HMPPS applying different description summarization strategies on Office Products dataset. Here, $\text{WC}_p$ denotes the description word count of one product and $\text{WC}_h$ denotes the user history word count of one sample composed of multiple products.} 
  \renewcommand{\arraystretch}{1.1}
  \begin{tabular}{c|cc|cc|cc}
    \toprule
    \multirow{2}{*}{Sum Prompt} & \multicolumn{2}{c|}{$\text{WC}_p$}& \multicolumn{2}{c|}{$\text{WC}_h$} & \multirow{2}{*}{N@4} & \multirow{2}{*}{R@4} \\
    \cmidrule{2-5}
    ~  & Avg & Max  & Avg  & Max  & ~ & ~ \\
    \midrule  
    None  & 106 & 3538 & 651 & 5507 & 0.401 & 0.524 \\
    \midrule
    Direct   & 38 & \textbf{88} & 281 & \textbf{557} & 0.401 & 0.525 \\
    Reasoning-based  & \textbf{34} & 1513 & \textbf{271} & 2103 & 0.400 & 0.522 \\
    Perspective-guided   & 41 & 513 & 287 & 932 &\textbf{0.403} &\textbf{0.529} \\
    \bottomrule
  \end{tabular}  
  \label{tab:description_summarization}  
  \end{center} 
  \vspace{-0.3cm}
  \end{table} 
  
  \begin{figure}[t] 
    \centering 
    \includegraphics[width=0.48\textwidth]{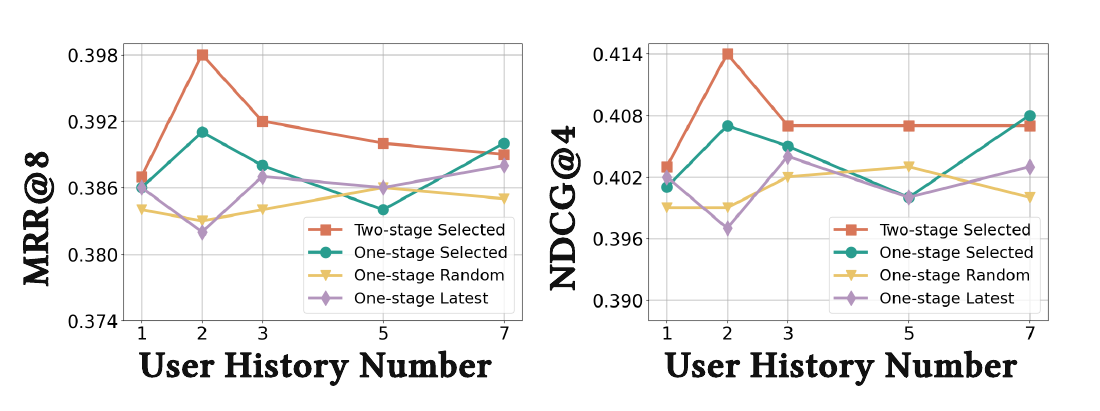}
    \caption{Ablation results of HMPPS with different types of user history on Office Products dataset.}
  \vspace{-0.2cm}
  \label{fig:user_history}
  \vspace{-0.3cm}
  \end{figure}

\subsubsection{User History Selection by Two-stage Training Paradigm.}
To demonstrate the effectiveness of the two-stage training paradigm for user history selection, we compare the search results of four variants:
1) \textbf{One-stage Latest} takes the latest purchased products as user history input to train HMPPS for only one stage;
2) \textbf{One-stage Random} randomly selects purchased products as user history;   
3) \textbf{One-stage Selected} utilizes the finetuned MLLM of One-stage Random to select user history related to the query and candidate product and then train HMPPS based on the selected user history only for one stage, similar to the previous two variants;
and 4) \textbf{Two-stage Selected} applies the selected user history extracted by the finetuned MLLM of One-stage Random and further trains the MLLM for the second stage of HMPPS.

From the experimental results shown in Figure ~\ref{fig:user_history}, we can obtain the following observations:  
1) even though without the first-stage learned parameters, training HMPPS with only two selected historical products can surpass both the best results of five random historical products and nine latest historical products. This demonstrates that the finetuned MLLM is effective for capturing exact user history relevant to query and candidate product. It not only enhances the PPS performance but also reduces the computation burden of the second-stage training, where the inference speed quadruples with decreased history size;
and 2) further training HMPPS with two selected historical products based on the first-stage learned parameters achieves the best performance. This not only verifies the history selection accuracy of the finetuned MLLM but also the effectiveness of the two-stage training paradigm in enhancing the robustness of HMPPS, due to valid user history selection, to some extent, can be regarded as data augmentation.

\begin{figure}[t]
  \centering 
  \includegraphics[width=0.48\textwidth]{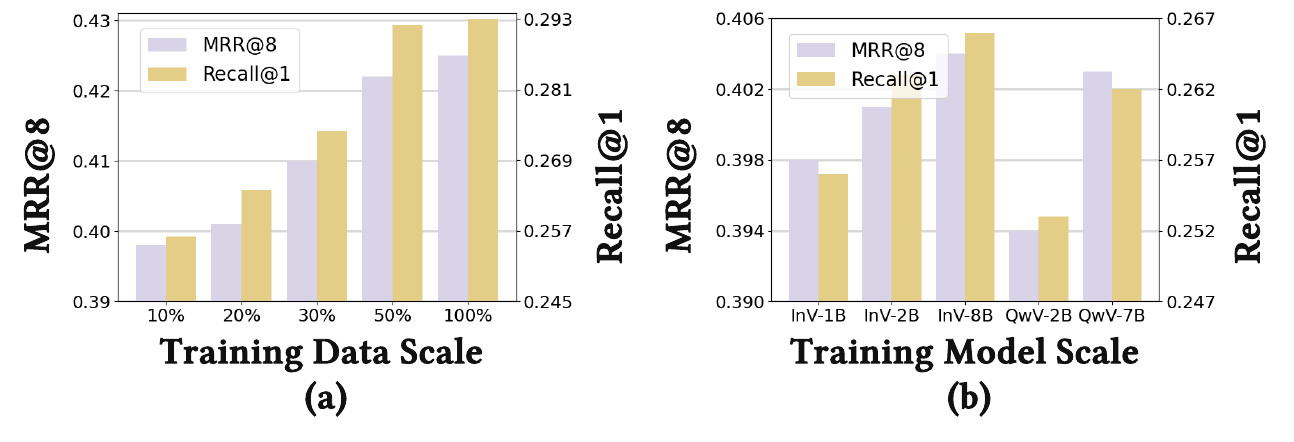}
  \vspace{-0.7cm}
  \caption{Ablation results of HMPPS trained on different scales of training data and models on Office Products dataset. Here, InV and QwV refer to MLLMs of InternVL2 and Qwen2VL.}
\label{fig:training_scale}
\vspace{-0.5cm}
\end{figure}

\begin{figure}[t] 
  \centering 
  \includegraphics[width=0.47\textwidth]{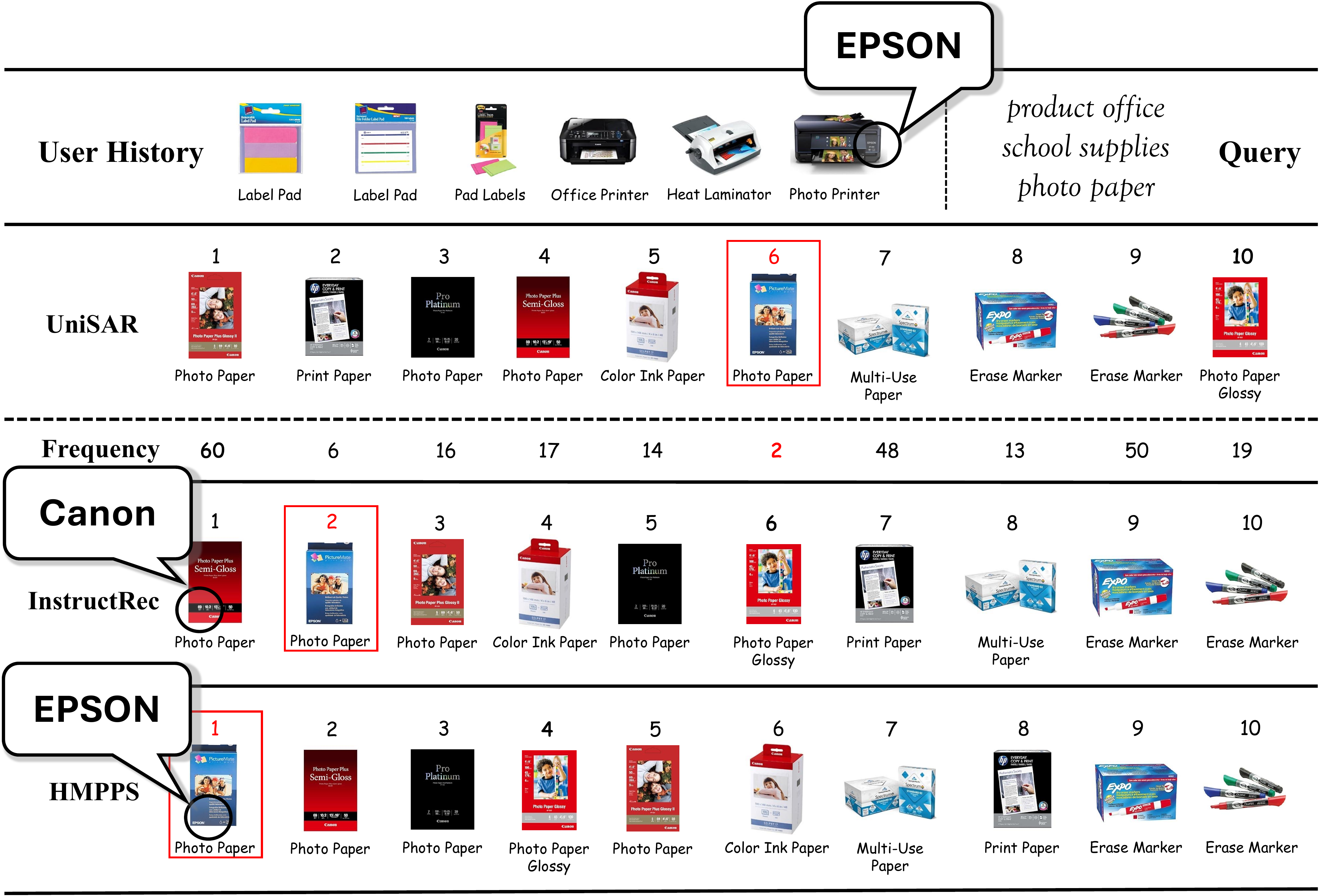}
  \caption{Example PPS results of UniSAR employing HMPPS \emph{vs.} InstructRec to rerank its search results on Office Products dataset. Here, $\text{Frequency}$ denotes the occurance frequency of the corresponding product in the training dataset and the target product is labeled by a red box.}
\label{fig:case_study}
\vspace{-0.3cm}
\end{figure} 
 
\subsection{Training Scale (RQ4)} 
\subsubsection{Training Data Scale.}
To further explore the potential of HMPPS, we experiment on different data scales for HMPPS training, which are 10\%, 20\%, 30\%, 50\% and 100\%. 
From the experimental results in Figure ~\ref{fig:training_scale} (a), we can observe that HMPPS can definitely perform better by training on more data for almost all metrics. There exists significant improvement from using 10\% to 100\% training data, which reveals the enormous potential of HMPPS for boosting PPS.

\subsubsection{Training Model Scale.}
\label{sec:mllm_scale}
One important practice of migrating MLLM to downstream tasks is to implement the proposed approaches with MLLM of different model scales.
Larger-scale MLLM always stands for more powerful understanding, reasoning and generation capabilities that contribute to performance improvement.  
To explore the potential of HMPPS with MLLMs of different model scales, we experiment on applying different scales of InternVL2 as the MLLM backbone of HMPPS.
We additionally take Qwen2VL~\cite{Qwen2VL} as the HMPPS MLLM backbone to validate the university of HMPPS.
The evaluation result is shown in Figure ~\ref{fig:training_scale} (b), which uncovers that larger-scale MLLMs, no matter which types they are, can indeed lead to improvement for HMPPS due to their remarkable progress in multimodal content understanding and reasoning with more well-learned parameters.

\vspace{-0.2cm}
\subsection{Case Study (RQ5)}
We provide case study in Figure ~\ref{fig:case_study} to qualitatively illustrate the advantage of HMPPS. 
To demonstrate that HMPPS can effectively make up for the limitations of ID-based approaches and outperforms LLM-based methods in content understanding, we compare HMPPS results with those of ID-based UniSAR and LLM-based InstructRec and we can observe that:
1) UniSAR ranks the most frequent product as the first while the target one with the lowest frequency are ranked sixth. However, HMPPS can successfully capture the target product which validates that HMPPS can relieve the data bias problem of ID-based methods;
and 2) for hard cases like similar candidates, HMPPS performs better than InstructRec since it can find out fine-grained differences according to multimodal contents, \emph{e.g.,} brand information absent in textual contents but showed in visual figures.

\subsection{Online Evaluation (RQ6)}
To demonstrate the practicability of HMPPS, we conduct A/B testing on an online search system that boasts billion-level daily active users.
This specific practice involves the following three procedures:
1) we firstly train HMPPS based on the offline data with a powerful large-scale MLLM;
2) the trained model is then distilled into a small-scale variant for online application;
and 3) since the online system is composed of multiple processing modules, we utilize the distilled model to predict the search probability, serving as one of the factors of the final search result.

During A/B testing, we replace the search probability predicted by a conventional multimodal transformer, which resembles Bert4Rec~\cite{sun2019bert4rec}, with that extracted by HMPPS.
The online experiment, conducted over a span of 14 days, yields a 0.53\% gain for query-ctr and a 0.77\% increase in efficient click count with p-value = 1.16\%, which demonstrates a significant improvement for highly-optimized real-world systems.
Here, query-ctr assesses whether users click on the items returned by the online system in response to their search queries. Meanwhile, the efficient click count quantifies the number of items that users have viewed for more than 5 seconds.

The inference time of ranking 10 candidate items for a query is 22 microseconds in average.
Since HMPPS is targeted for reranking, which just ranks a handful of retrieved candidates for more accurate and finegrained search result, its latency and computation expense can be acceptable for online application.

\section{Conclusion}
In this paper, to address the limitations of LLM-based approaches in PPS reranking, we proposed a novel method, HMPPS, harnessing pre-trained MLLMs to deal with PPS based on multimodal contents.
Except for adapting MLLM to PPS by converting the search task into a multimodal language understanding problem, we designed two query-aware refinement modules to reduce the redundancy in PPS input, which is a perspective-guided summarization module for product description refinement and a two-stage training paradigm for user history selection. Both of these two modules improve the prediction accuracy and reduce the computation cost of HMPPS.
Extensive experiments were conducted on four datasets to demonstrate the effectiveness of HMPPS and the evident gain in online A/B testing also validated the practicability of HMPPS.

\begin{acks}
This work is supported by the National Science Foundation of China (62072232), the Collaborative Innovation Center of Novel Software Technology and Industrialization, and the Young Elite Scientists Sponsorship Program by CAST (2023QNRC001).
\end{acks}

\bibliographystyle{ACM-Reference-Format}
\balance
\bibliography{sample-base}


\begin{thebibliography}{44}

%%% the Web address.

\ifx \showCODEN    \undefined \def \showCODEN     #1{\unskip}     \fi
\ifx \showISBNx    \undefined \def \showISBNx     #1{\unskip}     \fi
\ifx \showISBNxiii \undefined \def \showISBNxiii  #1{\unskip}     \fi
\ifx \showISSN     \undefined \def \showISSN      #1{\unskip}     \fi
\ifx \showLCCN     \undefined \def \showLCCN      #1{\unskip}     \fi
\ifx \shownote     \undefined \def \shownote      #1{#1}          \fi
\ifx \showarticletitle \undefined \def \showarticletitle #1{#1}   \fi
\ifx \showURL      \undefined \def \showURL       {\relax}        \fi
% invisible to TeX
\providecommand\bibfield[2]{#2}
\providecommand\bibinfo[2]{#2}
\providecommand\natexlab[1]{#1}
\providecommand\showeprint[2][]{arXiv:#2}

\bibitem[Ai et~al\mbox{.}(2019a)]%
        {ai2019zero}
\bibfield{author}{\bibinfo{person}{Qingyao Ai}, \bibinfo{person}{Daniel~N Hill}, \bibinfo{person}{SVN Vishwanathan}, {and} \bibinfo{person}{W~Bruce Croft}.} \bibinfo{year}{2019}\natexlab{a}.
\newblock \showarticletitle{A zero attention model for personalized product search}. In \bibinfo{booktitle}{\emph{Proceedings of the ACM International Conference on Information and Knowledge Management}}. \bibinfo{pages}{379--388}.
\newblock


\bibitem[Ai and Narayanan.~R(2021)]%
        {ai2021model}
\bibfield{author}{\bibinfo{person}{Qingyao Ai} {and} \bibinfo{person}{Lakshmi Narayanan.~R}.} \bibinfo{year}{2021}\natexlab{}.
\newblock \showarticletitle{Model-agnostic vs. model-intrinsic interpretability for explainable product search}. In \bibinfo{booktitle}{\emph{Proceedings of the 30th ACM International Conference on Information \& Knowledge Management}}. \bibinfo{pages}{5--15}.
\newblock


\bibitem[Ai et~al\mbox{.}(2017)]%
        {ai2017learning}
\bibfield{author}{\bibinfo{person}{Qingyao Ai}, \bibinfo{person}{Yongfeng Zhang}, \bibinfo{person}{Keping Bi}, \bibinfo{person}{Xu Chen}, {and} \bibinfo{person}{W~Bruce Croft}.} \bibinfo{year}{2017}\natexlab{}.
\newblock \showarticletitle{Learning a hierarchical embedding model for personalized product search}. In \bibinfo{booktitle}{\emph{Proceedings of the International ACM SIGIR Conference on Research and Development in Information Retrieval}}. \bibinfo{pages}{645--654}.
\newblock


\bibitem[Ai et~al\mbox{.}(2019b)]%
        {ai2019explainable}
\bibfield{author}{\bibinfo{person}{Qingyao Ai}, \bibinfo{person}{Yongfeng Zhang}, \bibinfo{person}{Keping Bi}, {and} \bibinfo{person}{W~Bruce Croft}.} \bibinfo{year}{2019}\natexlab{b}.
\newblock \showarticletitle{Explainable product search with a dynamic relation embedding model}. In \bibinfo{booktitle}{\emph{ACM Transactions on Information Systems}}, Vol.~\bibinfo{volume}{38}. \bibinfo{pages}{1--29}.
\newblock


\bibitem[Bi et~al\mbox{.}(2020)]%
        {bi2020transformer}
\bibfield{author}{\bibinfo{person}{Keping Bi}, \bibinfo{person}{Qingyao Ai}, {and} \bibinfo{person}{W~Bruce Croft}.} \bibinfo{year}{2020}\natexlab{}.
\newblock \showarticletitle{A transformer-based embedding model for personalized product search}. In \bibinfo{booktitle}{\emph{Proceedings of the International ACM SIGIR Conference on Research and Development in Information Retrieval}}. \bibinfo{pages}{1521--1524}.
\newblock


\bibitem[Bi et~al\mbox{.}(2021)]%
        {bi2021learning}
\bibfield{author}{\bibinfo{person}{Keping Bi}, \bibinfo{person}{Qingyao Ai}, {and} \bibinfo{person}{W~Bruce Croft}.} \bibinfo{year}{2021}\natexlab{}.
\newblock \showarticletitle{Learning a fine-grained review-based transformer model for personalized product search}. In \bibinfo{booktitle}{\emph{Proceedings of the International ACM SIGIR Conference on Research and Development in Information Retrieval}}. \bibinfo{pages}{123--132}.
\newblock


\bibitem[Brinkmann et~al\mbox{.}(2023)]%
        {brinkmann2023product}
\bibfield{author}{\bibinfo{person}{Alexander Brinkmann}, \bibinfo{person}{Roee Shraga}, \bibinfo{person}{Reng~Chiz Der}, {and} \bibinfo{person}{Christian Bizer}.} \bibinfo{year}{2023}\natexlab{}.
\newblock \showarticletitle{Product Information Extraction using ChatGPT}. In \bibinfo{booktitle}{\emph{arXiv preprint arXiv:2306.14921}}.
\newblock


\bibitem[Chen et~al\mbox{.}(2023)]%
        {chen2023bias}
\bibfield{author}{\bibinfo{person}{Jiawei Chen}, \bibinfo{person}{Hande Dong}, \bibinfo{person}{Xiang Wang}, \bibinfo{person}{Fuli Feng}, \bibinfo{person}{Meng Wang}, {and} \bibinfo{person}{Xiangnan He}.} \bibinfo{year}{2023}\natexlab{}.
\newblock \showarticletitle{Bias and debias in recommender system: A survey and future directions}. In \bibinfo{booktitle}{\emph{ACM Transactions on Information Systems}}, Vol.~\bibinfo{volume}{41}. \bibinfo{pages}{1--39}.
\newblock


\bibitem[Chen et~al\mbox{.}(2024)]%
        {chen2024internvl}
\bibfield{author}{\bibinfo{person}{Zhe Chen}, \bibinfo{person}{Jiannan Wu}, \bibinfo{person}{Wenhai Wang}, \bibinfo{person}{Weijie Su}, \bibinfo{person}{Guo Chen}, \bibinfo{person}{Sen Xing}, \bibinfo{person}{Muyan Zhong}, \bibinfo{person}{Qinglong Zhang}, \bibinfo{person}{Xizhou Zhu}, \bibinfo{person}{Lewei Lu}, {et~al\mbox{.}}} \bibinfo{year}{2024}\natexlab{}.
\newblock \showarticletitle{Internvl: Scaling up vision foundation models and aligning for generic visual-linguistic tasks}. In \bibinfo{booktitle}{\emph{Proceedings of the IEEE/CVF Conference on Computer Vision and Pattern Recognition}}. \bibinfo{pages}{24185--24198}.
\newblock


\bibitem[Cheng et~al\mbox{.}(2022)]%
        {cheng2022ihgnn}
\bibfield{author}{\bibinfo{person}{Dian Cheng}, \bibinfo{person}{Jiawei Chen}, \bibinfo{person}{Wenjun Peng}, \bibinfo{person}{Wenqin Ye}, \bibinfo{person}{Fuyu Lv}, \bibinfo{person}{Tao Zhuang}, \bibinfo{person}{Xiaoyi Zeng}, {and} \bibinfo{person}{Xiangnan He}.} \bibinfo{year}{2022}\natexlab{}.
\newblock \showarticletitle{Ihgnn: Interactive hypergraph neural network for personalized product search}. In \bibinfo{booktitle}{\emph{Proceedings of the ACM Web Conference}}. \bibinfo{pages}{256--265}.
\newblock


\bibitem[Dai et~al\mbox{.}(2023)]%
        {dai2023contrastive}
\bibfield{author}{\bibinfo{person}{Shitong Dai}, \bibinfo{person}{Jiongnan Liu}, \bibinfo{person}{Zhicheng Dou}, \bibinfo{person}{Haonan Wang}, \bibinfo{person}{Lin Liu}, \bibinfo{person}{Bo Long}, {and} \bibinfo{person}{Ji-Rong Wen}.} \bibinfo{year}{2023}\natexlab{}.
\newblock \showarticletitle{Contrastive Learning for User Sequence Representation in Personalized Product Search}. In \bibinfo{booktitle}{\emph{Proceedings of the ACM SIGKDD Conference on Knowledge Discovery and Data Mining}}. \bibinfo{pages}{380--389}.
\newblock


\bibitem[Fan et~al\mbox{.}(2022)]%
        {fan2022modeling}
\bibfield{author}{\bibinfo{person}{Lu Fan}, \bibinfo{person}{Qimai Li}, \bibinfo{person}{Bo Liu}, \bibinfo{person}{Xiao-Ming Wu}, \bibinfo{person}{Xiaotong Zhang}, \bibinfo{person}{Fuyu Lv}, \bibinfo{person}{Guli Lin}, \bibinfo{person}{Sen Li}, \bibinfo{person}{Taiwei Jin}, {and} \bibinfo{person}{Keping Yang}.} \bibinfo{year}{2022}\natexlab{}.
\newblock \showarticletitle{Modeling user behavior with graph convolution for personalized product search}. In \bibinfo{booktitle}{\emph{Proceedings of the ACM Web Conference}}. \bibinfo{pages}{203--212}.
\newblock


\bibitem[Guo et~al\mbox{.}(2023)]%
        {guo2023query}
\bibfield{author}{\bibinfo{person}{Tong Guo}, \bibinfo{person}{Xuanping Li}, \bibinfo{person}{Haitao Yang}, \bibinfo{person}{Xiao Liang}, \bibinfo{person}{Yong Yuan}, \bibinfo{person}{Jingyou Hou}, \bibinfo{person}{Bingqing Ke}, \bibinfo{person}{Chao Zhang}, \bibinfo{person}{Junlin He}, \bibinfo{person}{Shunyu Zhang}, {et~al\mbox{.}}} \bibinfo{year}{2023}\natexlab{}.
\newblock \showarticletitle{Query-dominant User Interest Network for Large-Scale Search Ranking}. In \bibinfo{booktitle}{\emph{Proceedings of the ACM International Conference on Information and Knowledge Management}}. \bibinfo{pages}{629--638}.
\newblock


\bibitem[Guo et~al\mbox{.}({[n.\,d.]})]%
        {guo2019attentive}
\bibfield{author}{\bibinfo{person}{Yangyang Guo}, \bibinfo{person}{Zhiyong Cheng}, \bibinfo{person}{Liqiang Nie}, \bibinfo{person}{Yinglong Wang}, \bibinfo{person}{Jun Ma}, {and} \bibinfo{person}{Mohan Kankanhalli}.} \bibinfo{year}{[n.\,d.]}\natexlab{}.
\newblock \showarticletitle{Attentive long short-term preference modeling for personalized product search}. In \bibinfo{booktitle}{\emph{ACM Transactions on Information Systems}}, Vol.~\bibinfo{volume}{37}. \bibinfo{pages}{1--27}.
\newblock


\bibitem[Guo et~al\mbox{.}(2018)]%
        {guo2018multi}
\bibfield{author}{\bibinfo{person}{Yangyang Guo}, \bibinfo{person}{Zhiyong Cheng}, \bibinfo{person}{Liqiang Nie}, \bibinfo{person}{Xin-Shun Xu}, {and} \bibinfo{person}{Mohan Kankanhalli}.} \bibinfo{year}{2018}\natexlab{}.
\newblock \showarticletitle{Multi-modal preference modeling for product search}. In \bibinfo{booktitle}{\emph{Proceedings of the ACM international conference on Multimedia}}. \bibinfo{pages}{1865--1873}.
\newblock


\bibitem[Jiang et~al\mbox{.}(2020)]%
        {jiang2020end}
\bibfield{author}{\bibinfo{person}{Jyun-Yu Jiang}, \bibinfo{person}{Tao Wu}, \bibinfo{person}{Georgios Roumpos}, \bibinfo{person}{Heng-Tze Cheng}, \bibinfo{person}{Xinyang Yi}, \bibinfo{person}{Ed Chi}, \bibinfo{person}{Harish Ganapathy}, \bibinfo{person}{Nitin Jindal}, \bibinfo{person}{Pei Cao}, {and} \bibinfo{person}{Wei Wang}.} \bibinfo{year}{2020}\natexlab{}.
\newblock \showarticletitle{End-to-end deep attentive personalized item retrieval for online content-sharing platforms}. In \bibinfo{booktitle}{\emph{Proceedings of the ACM Web Conference}}. \bibinfo{pages}{2870--2877}.
\newblock


\bibitem[Liu et~al\mbox{.}(2022)]%
        {liu2022category}
\bibfield{author}{\bibinfo{person}{Jiongnan Liu}, \bibinfo{person}{Zhicheng Dou}, \bibinfo{person}{Qiannan Zhu}, {and} \bibinfo{person}{Ji-Rong Wen}.} \bibinfo{year}{2022}\natexlab{}.
\newblock \showarticletitle{A category-aware multi-interest model for personalized product search}. In \bibinfo{booktitle}{\emph{Proceedings of the ACM Web Conference}}. \bibinfo{pages}{360--368}.
\newblock


\bibitem[Liu et~al\mbox{.}(2020)]%
        {liu2020structural}
\bibfield{author}{\bibinfo{person}{Shang Liu}, \bibinfo{person}{Wanli Gu}, \bibinfo{person}{Gao Cong}, {and} \bibinfo{person}{Fuzheng Zhang}.} \bibinfo{year}{2020}\natexlab{}.
\newblock \showarticletitle{Structural relationship representation learning with graph embedding for personalized product search}. In \bibinfo{booktitle}{\emph{Proceedings of the ACM International Conference on Information and Knowledge Management}}. \bibinfo{pages}{915--924}.
\newblock


\bibitem[Ma et~al\mbox{.}(2024)]%
        {ma2024finetuning}
\bibfield{author}{\bibinfo{person}{Xueguang Ma}, \bibinfo{person}{Liang Wang}, \bibinfo{person}{Nan Yang}, \bibinfo{person}{Furu Wei}, {and} \bibinfo{person}{Jimmy Lin}.} \bibinfo{year}{2024}\natexlab{}.
\newblock \showarticletitle{Fine-Tuning LLaMA for Multi-Stage Text Retrieval}. In \bibinfo{booktitle}{\emph{Proceedings of the International ACM SIGIR Conference on Research and Development in Information Retrieval}}. \bibinfo{pages}{2421--2425}.
\newblock


\bibitem[McAuley et~al\mbox{.}(2015)]%
        {mcauley2015inferring}
\bibfield{author}{\bibinfo{person}{Julian McAuley}, \bibinfo{person}{Rahul Pandey}, {and} \bibinfo{person}{Jure Leskovec}.} \bibinfo{year}{2015}\natexlab{}.
\newblock \showarticletitle{Inferring networks of substitutable and complementary products}. In \bibinfo{booktitle}{\emph{Proceedings of the ACM SIGKDD Conference on Knowledge Discovery and Data Mining}}. \bibinfo{pages}{785--794}.
\newblock


\bibitem[Pan et~al\mbox{.}(2021)]%
        {pan2021personalized}
\bibfield{author}{\bibinfo{person}{Yaoxin Pan}, \bibinfo{person}{Shangsong Liang}, \bibinfo{person}{Jiaxin Ren}, \bibinfo{person}{Zaiqiao Meng}, {and} \bibinfo{person}{Qiang Zhang}.} \bibinfo{year}{2021}\natexlab{}.
\newblock \showarticletitle{Personalized, sequential, attentive, metric-aware product search}.
\newblock \bibinfo{journal}{\emph{ACM Transactions on Information Systems}} \bibinfo{volume}{40}, \bibinfo{number}{2}, \bibinfo{pages}{1--29}.
\newblock


\bibitem[Pi et~al\mbox{.}(2019)]%
        {pi2019practice}
\bibfield{author}{\bibinfo{person}{Qi Pi}, \bibinfo{person}{Weijie Bian}, \bibinfo{person}{Guorui Zhou}, \bibinfo{person}{Xiaoqiang Zhu}, {and} \bibinfo{person}{Kun Gai}.} \bibinfo{year}{2019}\natexlab{}.
\newblock \showarticletitle{Practice on long sequential user behavior modeling for click-through rate prediction}. In \bibinfo{booktitle}{\emph{Proceedings of the ACM SIGKDD Conference on Knowledge Discovery and Data Mining}}. \bibinfo{pages}{2671--2679}.
\newblock


\bibitem[Pi et~al\mbox{.}(2020)]%
        {pi2020search}
\bibfield{author}{\bibinfo{person}{Qi Pi}, \bibinfo{person}{Guorui Zhou}, \bibinfo{person}{Yujing Zhang}, \bibinfo{person}{Zhe Wang}, \bibinfo{person}{Lejian Ren}, \bibinfo{person}{Ying Fan}, \bibinfo{person}{Xiaoqiang Zhu}, {and} \bibinfo{person}{Kun Gai}.} \bibinfo{year}{2020}\natexlab{}.
\newblock \showarticletitle{Search-based user interest modeling with lifelong sequential behavior data for click-through rate prediction}. In \bibinfo{booktitle}{\emph{Proceedings of the ACM International Conference on Information and Knowledge Management}}. \bibinfo{pages}{2685--2692}.
\newblock


\bibitem[Ren et~al\mbox{.}(2019)]%
        {ren2019lifelong}
\bibfield{author}{\bibinfo{person}{Kan Ren}, \bibinfo{person}{Jiarui Qin}, \bibinfo{person}{Yuchen Fang}, \bibinfo{person}{Weinan Zhang}, \bibinfo{person}{Lei Zheng}, \bibinfo{person}{Weijie Bian}, \bibinfo{person}{Guorui Zhou}, \bibinfo{person}{Jian Xu}, \bibinfo{person}{Yong Yu}, \bibinfo{person}{Xiaoqiang Zhu}, {et~al\mbox{.}}} \bibinfo{year}{2019}\natexlab{}.
\newblock \showarticletitle{Lifelong sequential modeling with personalized memorization for user response prediction}. In \bibinfo{booktitle}{\emph{Proceedings of the International ACM SIGIR Conference on Research and Development in Information Retrieval}}. \bibinfo{pages}{565--574}.
\newblock


\bibitem[Ren et~al\mbox{.}(2024)]%
        {ren2024representation}
\bibfield{author}{\bibinfo{person}{Xubin Ren}, \bibinfo{person}{Wei Wei}, \bibinfo{person}{Lianghao Xia}, \bibinfo{person}{Lixin Su}, \bibinfo{person}{Suqi Cheng}, \bibinfo{person}{Junfeng Wang}, \bibinfo{person}{Dawei Yin}, {and} \bibinfo{person}{Chao Huang}.} \bibinfo{year}{2024}\natexlab{}.
\newblock \showarticletitle{Representation learning with large language models for recommendation}. In \bibinfo{booktitle}{\emph{Proceedings of the ACM Web Conference}}. \bibinfo{pages}{3464--3475}.
\newblock


\bibitem[Shen et~al\mbox{.}(2024)]%
        {shen2024seminar}
\bibfield{author}{\bibinfo{person}{Kaiming Shen}, \bibinfo{person}{Xichen Ding}, \bibinfo{person}{Zixiang Zheng}, \bibinfo{person}{Yuqi Gong}, \bibinfo{person}{Qianqian Li}, \bibinfo{person}{Zhongyi Liu}, {and} \bibinfo{person}{Guannan Zhang}.} \bibinfo{year}{2024}\natexlab{}.
\newblock \showarticletitle{SEMINAR: Search Enhanced Multi-modal Interest Network and Approximate Retrieval for Lifelong Sequential Recommendation}. In \bibinfo{booktitle}{\emph{arXiv preprint arXiv:2407.10714}}.
\newblock


\bibitem[Shen et~al\mbox{.}(2022)]%
        {shen2022hierarchically}
\bibfield{author}{\bibinfo{person}{Qijie Shen}, \bibinfo{person}{Hong Wen}, \bibinfo{person}{Jing Zhang}, {and} \bibinfo{person}{Qi Rao}.} \bibinfo{year}{2022}\natexlab{}.
\newblock \showarticletitle{Hierarchically fusing long and short-term user interests for click-through rate prediction in product search}. In \bibinfo{booktitle}{\emph{Proceedings of the ACM International Conference on Information and Knowledge Management}}. \bibinfo{pages}{1767--1776}.
\newblock


\bibitem[Shi et~al\mbox{.}(2024)]%
        {shi2024unisar}
\bibfield{author}{\bibinfo{person}{Teng Shi}, \bibinfo{person}{Zihua Si}, \bibinfo{person}{Jun Xu}, \bibinfo{person}{Xiao Zhang}, \bibinfo{person}{Xiaoxue Zang}, \bibinfo{person}{Kai Zheng}, \bibinfo{person}{Dewei Leng}, \bibinfo{person}{Yanan Niu}, {and} \bibinfo{person}{Yang Song}.} \bibinfo{year}{2024}\natexlab{}.
\newblock \showarticletitle{UniSAR: Modeling User Transition Behaviors between Search and Recommendation}. In \bibinfo{booktitle}{\emph{Proceedings of the International ACM SIGIR Conference on Research and Development in Information Retrieval}}. \bibinfo{pages}{1029--1039}.
\newblock


\bibitem[Sun et~al\mbox{.}(2019)]%
        {sun2019bert4rec}
\bibfield{author}{\bibinfo{person}{Fei Sun}, \bibinfo{person}{Jun Liu}, \bibinfo{person}{Jian Wu}, \bibinfo{person}{Changhua Pei}, \bibinfo{person}{Xiao Lin}, \bibinfo{person}{Wenwu Ou}, {and} \bibinfo{person}{Peng Jiang}.} \bibinfo{year}{2019}\natexlab{}.
\newblock \showarticletitle{BERT4Rec: Sequential Recommendation with Bidirectional Encoder Representations from Transformer}. In \bibinfo{booktitle}{\emph{Proceedings of the ACM International Conference on Information and Knowledge Management}}. \bibinfo{pages}{1441--1450}.
\newblock


\bibitem[Thonet et~al\mbox{.}(2022)]%
        {thonet2022joint}
\bibfield{author}{\bibinfo{person}{Thibaut Thonet}, \bibinfo{person}{Jean-Michel Renders}, \bibinfo{person}{Mario Choi}, {and} \bibinfo{person}{Jinho Kim}.} \bibinfo{year}{2022}\natexlab{}.
\newblock \showarticletitle{Joint Personalized Search and Recommendation with Hypergraph Convolutional Networks}.
\newblock \bibinfo{journal}{\emph{Advances in Information Retrieval}}  \bibinfo{volume}{13185}, \bibinfo{pages}{443--456}.
\newblock


\bibitem[Torbati et~al\mbox{.}(2023)]%
        {torbati2023recommendations}
\bibfield{author}{\bibinfo{person}{Ghazaleh~Haratinezhad Torbati}, \bibinfo{person}{Anna Tigunova}, \bibinfo{person}{Andrew Yates}, {and} \bibinfo{person}{Gerhard Weikum}.} \bibinfo{year}{2023}\natexlab{}.
\newblock \showarticletitle{Recommendations by Concise User Profiles from Review Text}. In \bibinfo{booktitle}{\emph{arXiv preprint arXiv:2311.01314}}.
\newblock


\bibitem[Wang et~al\mbox{.}(2024)]%
        {Qwen2VL}
\bibfield{author}{\bibinfo{person}{Peng Wang}, \bibinfo{person}{Shuai Bai}, \bibinfo{person}{Sinan Tan}, \bibinfo{person}{Shijie Wang}, \bibinfo{person}{Zhihao Fan}, \bibinfo{person}{Jinze Bai}, \bibinfo{person}{Keqin Chen}, \bibinfo{person}{Xuejing Liu}, \bibinfo{person}{Jialin Wang}, \bibinfo{person}{Wenbin Ge}, \bibinfo{person}{Yang Fan}, \bibinfo{person}{Kai Dang}, \bibinfo{person}{Mengfei Du}, \bibinfo{person}{Xuancheng Ren}, \bibinfo{person}{Rui Men}, \bibinfo{person}{Dayiheng Liu}, \bibinfo{person}{Chang Zhou}, \bibinfo{person}{Jingren Zhou}, {and} \bibinfo{person}{Junyang Lin}.} \bibinfo{year}{2024}\natexlab{}.
\newblock \showarticletitle{Qwen2-VL: Enhancing Vision-Language Model's Perception of the World at Any Resolution}. In \bibinfo{booktitle}{\emph{arXiv preprint arXiv:2409.12191}}.
\newblock


\bibitem[Wei et~al\mbox{.}(2022)]%
        {wei2022chain}
\bibfield{author}{\bibinfo{person}{Jason Wei}, \bibinfo{person}{Xuezhi Wang}, \bibinfo{person}{Dale Schuurmans}, \bibinfo{person}{Maarten Bosma}, \bibinfo{person}{Fei Xia}, \bibinfo{person}{Ed Chi}, \bibinfo{person}{Quoc~V Le}, \bibinfo{person}{Denny Zhou}, {et~al\mbox{.}}} \bibinfo{year}{2022}\natexlab{}.
\newblock \showarticletitle{Chain-of-thought prompting elicits reasoning in large language models}. In \bibinfo{booktitle}{\emph{Advances in Neural Information Processing Systems}}, Vol.~\bibinfo{volume}{35}. \bibinfo{pages}{24824--24837}.
\newblock


\bibitem[Wei et~al\mbox{.}(2024)]%
        {wei2024towardsum}
\bibfield{author}{\bibinfo{person}{Tianxin Wei}, \bibinfo{person}{Bowen Jin}, \bibinfo{person}{Ruirui Li}, \bibinfo{person}{Hansi Zeng}, \bibinfo{person}{Zhengyang Wang}, \bibinfo{person}{Jianhui Sun}, \bibinfo{person}{Qingyu Yin}, \bibinfo{person}{Hanqing Lu}, \bibinfo{person}{Suhang Wang}, \bibinfo{person}{Jingrui He}, {and} \bibinfo{person}{Xianfeng Tang}.} \bibinfo{year}{2024}\natexlab{}.
\newblock \showarticletitle{Towards Unified Multi-Modal Personalization: Large Vision-Language Models for Generative Recommendation and Beyond}. In \bibinfo{booktitle}{\emph{Proceedings of the International Conference on Learning Representations}}. \bibinfo{pages}{1--19}.
\newblock


\bibitem[Wu et~al\mbox{.}(2023)]%
        {wu2023dynamic}
\bibfield{author}{\bibinfo{person}{Bin Wu}, \bibinfo{person}{Zaiqiao Meng}, {and} \bibinfo{person}{Shangsong Liang}.} \bibinfo{year}{2023}\natexlab{}.
\newblock \showarticletitle{Dynamic Bayesian Contrastive Predictive Coding Model for Personalized Product Search}. In \bibinfo{booktitle}{\emph{ACM Transactions on the Web}}, Vol.~\bibinfo{volume}{17}. \bibinfo{pages}{1--31}.
\newblock


\bibitem[Wu et~al\mbox{.}(2022)]%
        {wu2022meta}
\bibfield{author}{\bibinfo{person}{Bin Wu}, \bibinfo{person}{Zaiqiao Meng}, \bibinfo{person}{Qiang Zhang}, {and} \bibinfo{person}{Shangsong Liang}.} \bibinfo{year}{2022}\natexlab{}.
\newblock \showarticletitle{Meta-Learning Helps Personalized Product Search}. In \bibinfo{booktitle}{\emph{Proceedings of the ACM Web Conference}}. \bibinfo{pages}{2277--2287}.
\newblock


\bibitem[Wu et~al\mbox{.}(2025)]%
        {wu2025tfdcon}
\bibfield{author}{\bibinfo{person}{Jiahao Wu}, \bibinfo{person}{Qijiong Liu}, \bibinfo{person}{Hengchang Hu}, \bibinfo{person}{Wenqi Fan}, \bibinfo{person}{Shengcai Liu}, \bibinfo{person}{Qing Li}, \bibinfo{person}{Xiao-Ming Wu}, {and} \bibinfo{person}{Ke Tang}.} \bibinfo{year}{2025}\natexlab{}.
\newblock \showarticletitle{TF-DCon: Leveraging Large Language Models (LLMs) to Empower Training-Free Dataset Condensation for Content-Based Recommendation}. In \bibinfo{booktitle}{\emph{arXiv preprint arXiv:2310.09874}}.
\newblock


\bibitem[Xi et~al\mbox{.}(2024)]%
        {xi2024towards}
\bibfield{author}{\bibinfo{person}{Yunjia Xi}, \bibinfo{person}{Weiwen Liu}, \bibinfo{person}{Jianghao Lin}, \bibinfo{person}{Xiaoling Cai}, \bibinfo{person}{Hong Zhu}, \bibinfo{person}{Jieming Zhu}, \bibinfo{person}{Bo Chen}, \bibinfo{person}{Ruiming Tang}, \bibinfo{person}{Weinan Zhang}, {and} \bibinfo{person}{Yong Yu}.} \bibinfo{year}{2024}\natexlab{}.
\newblock \showarticletitle{Towards open-world recommendation with knowledge augmentation from large language models}. In \bibinfo{booktitle}{\emph{Proceedings of the ACM Conference on Recommender Systems}}. \bibinfo{pages}{12--22}.
\newblock


\bibitem[Xiao et~al\mbox{.}(2019)]%
        {xiao2019dynamic}
\bibfield{author}{\bibinfo{person}{Teng Xiao}, \bibinfo{person}{Jiaxin Ren}, \bibinfo{person}{Zaiqiao Meng}, \bibinfo{person}{Huan Sun}, {and} \bibinfo{person}{Shangsong Liang}.} \bibinfo{year}{2019}\natexlab{}.
\newblock \showarticletitle{Dynamic bayesian metric learning for personalized product search}. In \bibinfo{booktitle}{\emph{Proceedings of the ACM International Conference on Information and Knowledge Management}}. \bibinfo{pages}{1693--1702}.
\newblock


\bibitem[Yang et~al\mbox{.}(2024)]%
        {qwen2.5}
\bibfield{author}{\bibinfo{person}{An Yang}, \bibinfo{person}{Baosong Yang}, \bibinfo{person}{Beichen Zhang}, {and} \bibinfo{person}{et al}.} \bibinfo{year}{2024}\natexlab{}.
\newblock \showarticletitle{Qwen2.5 Technical Report}. In \bibinfo{booktitle}{\emph{arXiv preprint arXiv:2412.15115}}.
\newblock


\bibitem[Yuan et~al\mbox{.}(2023)]%
        {yuan2023go}
\bibfield{author}{\bibinfo{person}{Zheng Yuan}, \bibinfo{person}{Fajie Yuan}, \bibinfo{person}{Yu Song}, \bibinfo{person}{Youhua Li}, \bibinfo{person}{Junchen Fu}, \bibinfo{person}{Fei Yang}, \bibinfo{person}{Yunzhu Pan}, {and} \bibinfo{person}{Yongxin Ni}.} \bibinfo{year}{2023}\natexlab{}.
\newblock \showarticletitle{Where to go next for recommender systems? id-vs. modality-based recommender models revisited}. In \bibinfo{booktitle}{\emph{Proceedings of the International ACM SIGIR Conference on Research and Development in Information Retrieval}}. \bibinfo{pages}{2639--2649}.
\newblock


\bibitem[Zhang et~al\mbox{.}(2024)]%
        {zhang2023recommendation}
\bibfield{author}{\bibinfo{person}{Junjie Zhang}, \bibinfo{person}{Ruobing Xie}, \bibinfo{person}{Yupeng Hou}, \bibinfo{person}{Xin Zhao}, \bibinfo{person}{Leyu Lin}, {and} \bibinfo{person}{Ji-Rong Wen}.} \bibinfo{year}{2024}\natexlab{}.
\newblock \showarticletitle{Recommendation as instruction following: A large language model empowered recommendation approach}.
\newblock \bibinfo{journal}{\emph{ACM Transactions on Information Systems}} (\bibinfo{year}{2024}).
\newblock


\bibitem[Zhou et~al\mbox{.}(2019)]%
        {zhou2019deep}
\bibfield{author}{\bibinfo{person}{Guorui Zhou}, \bibinfo{person}{Na Mou}, \bibinfo{person}{Ying Fan}, \bibinfo{person}{Qi Pi}, \bibinfo{person}{Weijie Bian}, \bibinfo{person}{Chang Zhou}, \bibinfo{person}{Xiaoqiang Zhu}, {and} \bibinfo{person}{Kun Gai}.} \bibinfo{year}{2019}\natexlab{}.
\newblock \showarticletitle{Deep interest evolution network for click-through rate prediction}. In \bibinfo{booktitle}{\emph{Proceedings of the AAAI conference on Artificial Intelligence}}. \bibinfo{pages}{5941--5948}.
\newblock


\bibitem[Zhou et~al\mbox{.}(2018)]%
        {zhou2018deep}
\bibfield{author}{\bibinfo{person}{Guorui Zhou}, \bibinfo{person}{Xiaoqiang Zhu}, \bibinfo{person}{Chenru Song}, \bibinfo{person}{Ying Fan}, \bibinfo{person}{Han Zhu}, \bibinfo{person}{Xiao Ma}, \bibinfo{person}{Yanghui Yan}, \bibinfo{person}{Junqi Jin}, \bibinfo{person}{Han Li}, {and} \bibinfo{person}{Kun Gai}.} \bibinfo{year}{2018}\natexlab{}.
\newblock \showarticletitle{Deep interest network for click-through rate prediction}. In \bibinfo{booktitle}{\emph{Proceedings of the ACM SIGKDD Conference on Knowledge Discovery and Data Mining}}. \bibinfo{pages}{1059--1068}.
\newblock


\end{thebibliography}

\clearpage
\appendix

\twocolumn[{
\begin{center}
    \centering
    \vspace{-15pt}
    \includegraphics[width=\linewidth]{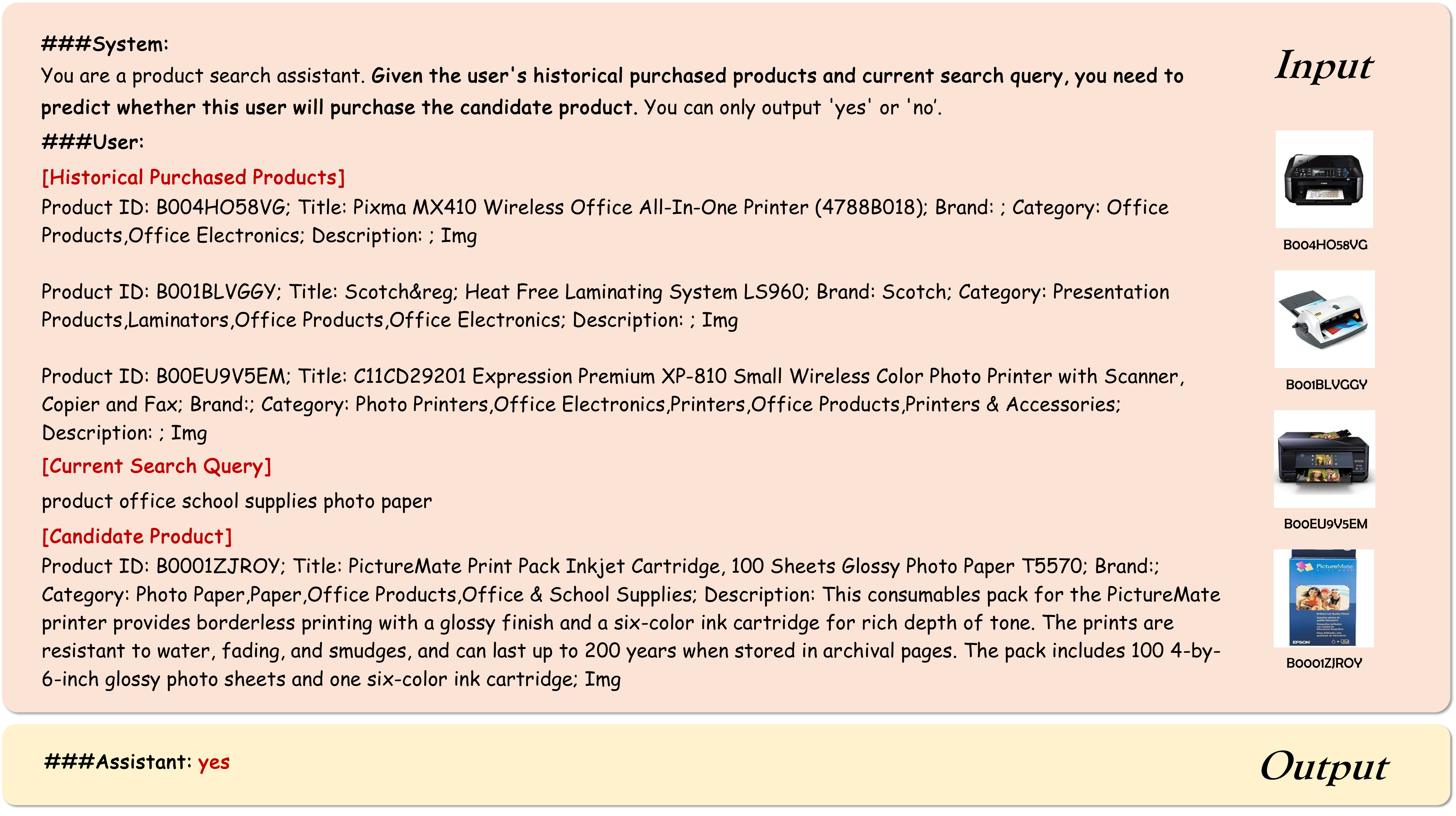}
    \setlength{\abovecaptionskip}{0mm}
    \captionof{figure}{\small \textbf{An example of MLLM-based personalized product search on Office Products dataset.}} 
    \vspace{15pt}
\label{fig:framework_example}
\end{center}}]

\section{Method Details}
\subsection{Example of HMPPS}

Figure ~\ref{fig:framework_example} showcases an example of the input and output of HMPPS, which applies MLLM to deal with PPS.
The task instruction at the beginning of the input prompt requires the MLLM to understand the relationship among user historical products, query and candidate product and output the search decision.
Each product involves multimodal contents where product ID, title, brand, category and description compose textual contents and visual displays make up for visual contents.
The output result demonstrates that HMPPS can generate precise search decision that relates to the user history and query via conducting analysis based on multimodal contents of query and product.

\subsection{Example of Perspective-guided Description Summarization}

\begin{figure*}[t]
  \centering 
  \includegraphics[width=0.95\textwidth]{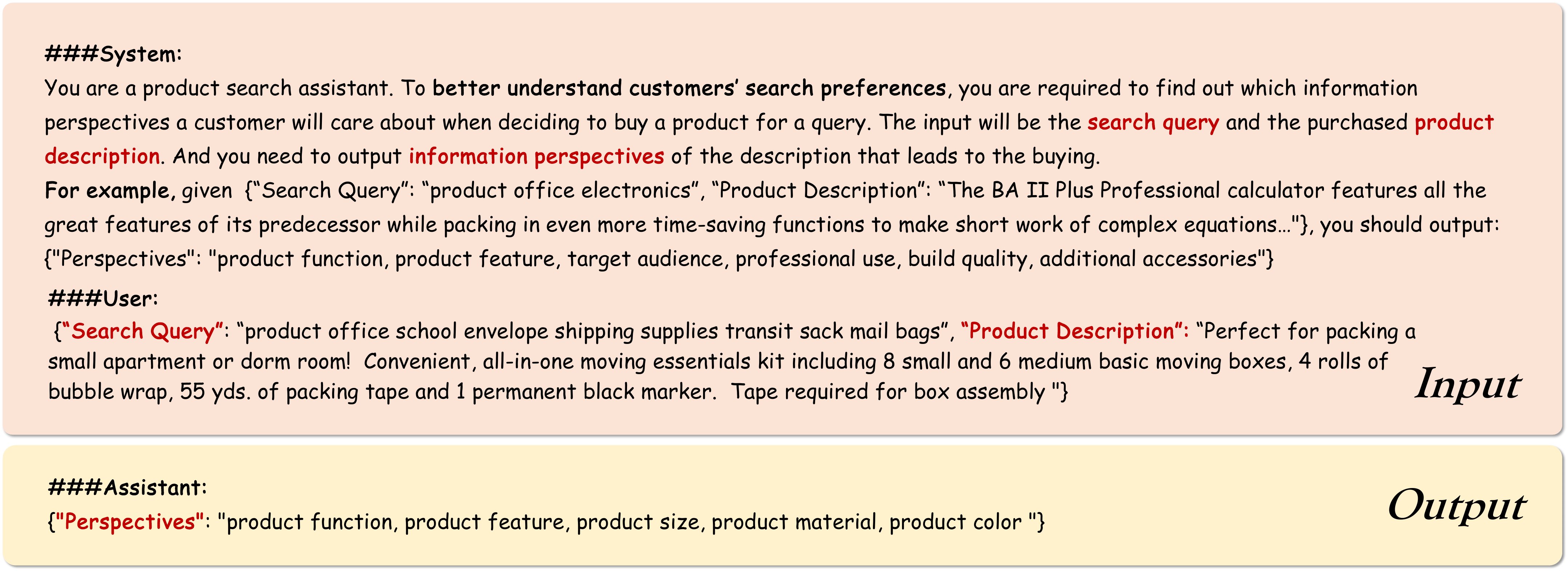}
  \caption{An example of perspective extraction in perspective-guided description summarization from Office Products dataset.}
\label{fig:perspective_sum_stage1_example}
\end{figure*} 

\begin{figure*}[t]
  \centering 
  \includegraphics[width=0.95\textwidth]{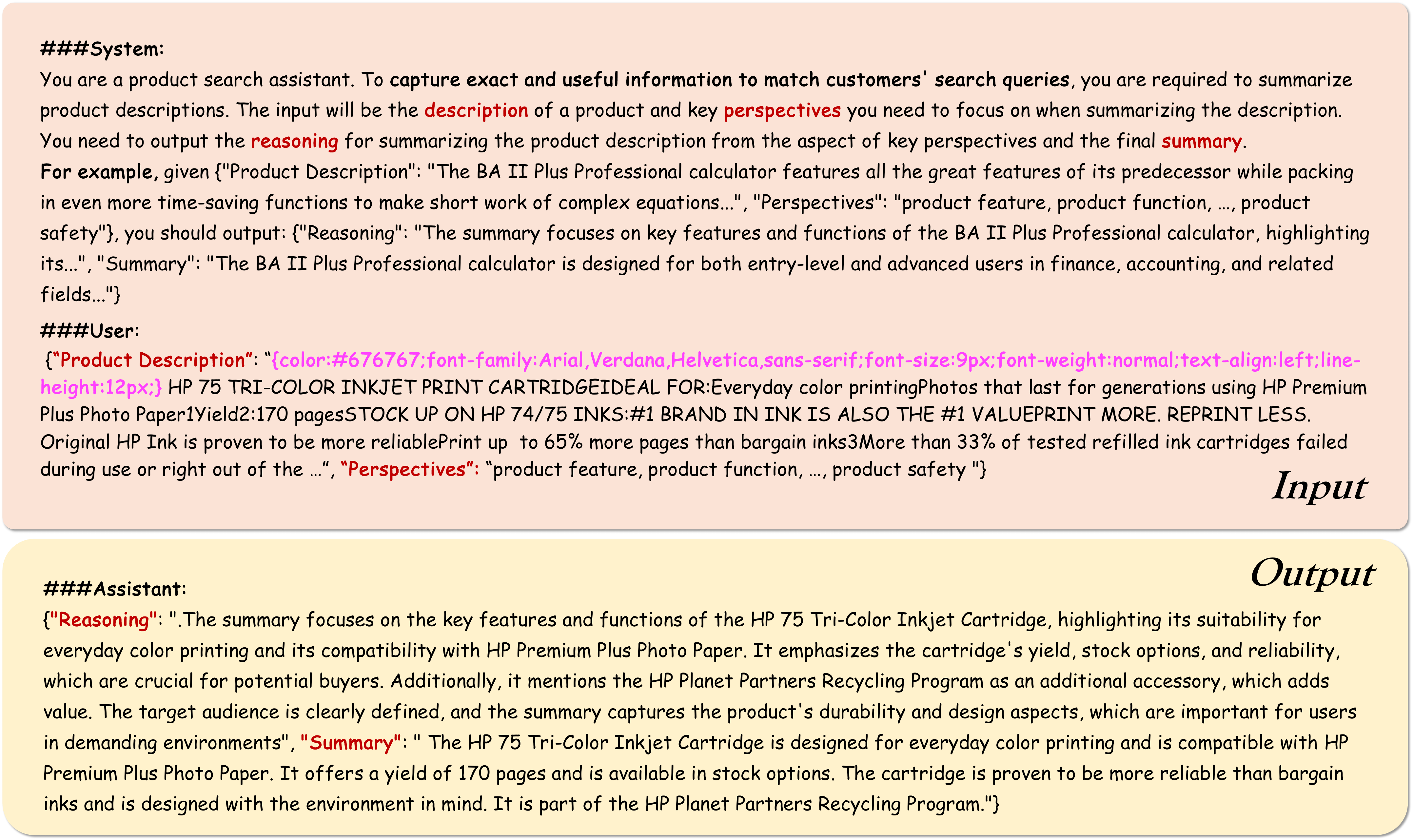}
  \caption{An example of summary generation in perspective-guided description summarization on Office Products dataset.}
\label{fig:perspective_sum_stage2_example}
\end{figure*}  
 
Figure ~\ref{fig:perspective_sum_stage1_example} showcases an example of the perspective extraction, which is the first step in perspective-guided description summarization module. 
As the instruction requires extracting information perspectives of product description with the guide of user query, it can capture perspectives that relates to customer search preferences.
Accurate perspectives are obtained with the help of the demonstration in prompt.
After collecting perspectives from all product descriptions, we retain a fixed number of perspectives as core information perspectives according to their frequency.  

Figure ~\ref{fig:perspective_sum_stage2_example} showcases an example of the summary generation, which is the second step in perspective-guided description summarization module. 
Core perspectives obtained in the previous step are leveraged to guide the process of summarization generation.
It can be observed that irrelevant noise, \emph{e.g.,} text font information, in the original description have been filtered out after the summarization.
Moreover, the generated summary not only includes precise product information but also meets the search preferences of customers as the summarization result is around the information perspectives that customers care about during product search.

\subsection{Online Search System Implementation}
\label{sec:online_implementation}
Due to the computation cost and inference efficiency of MLLM, it is impractical for an online search system to leverage a large-scale powerful MLLM to obtain the search result.
Therefore, we conduct the training process of HMPPS off-line with a large-scale MLLM of 72 billion parameters and then distill it to a small-scale variant of 300 million parameters.
Since there are many other valuable modules contributing to the large and complicated online search system, it is insufficient to merely depend on HMPPS to make the search decision.  
As a result, we leverage the distilled model to obtain the search probability, which serves as a component participating in the final search result fusion for online system.
It is noteworthy that we apply the proposed model only in reranking stage of the online search system that involves a handful of candidates filtered by the prepositive retrievers, for which the computation expense is acceptable.

To obtain a powerful search model, we collect 20 million samples from real user search logs of three months, covering 2 million active users and 8 million items, for HMPPS training.
To ensure that MLLM can effectively adapt to PPS domain, we additionally conduct specific data mining based on the relevance between user history and query to improve the data quality.

\section{Experimental Details}
\label{sec:experimental_details}
\subsection{Dataset details}
\label{sec:dataset_details}
We take 5-core Amazon product search dataset~\cite{mcauley2015inferring} as our experimental corpus, which contains product metadata and multimodal search logs from the Amazon website from May, 1996 to July, 2014.
To verify that HMPPS can adapt to diverse search scenarios, we select four typical subsets of the Amazon dataset which are \emph{Office Products} (Office), \emph{Cell Phones \& Accessories} (Cell), \emph{Beauty} and \emph{Sports \& Outdoors} (Sports). 
Statistics of these datasets are shown in Table ~\ref{tab:data_statistics}.
Following previous PPS works~\cite{ai2017learning, ai2019zero, ai2019explainable}, we split each dataset into train and test sets with a ratio of 7:3 and extract search queries using the same strategy as them.

\subsection{Evaluation Metrics.}
\label{sec:evaluation_metrics}
To evaluate the PPS performance of HMPPS, we utilize three typical search metrics which are Mean Reciprocal Rank (MRR), Normalized Discounted Cumulative Gain (NDCG) and Recall.
MRR and NDCG denotes search precision, which is calculated based on the position of positive products in the result list, evaluating ranking abilities of algorithms.
Recall corresponds to search coverage, concentrating on retrieval performance by calculating the ratio of positive products appearing in the returned products of search systems.
In this paper, we report MRR at position 8, NDCG at position {4} and Recall at positions \{4,1\}.
The higher values indicate better performance for all the metrics.

\begin{table}[t]
  \begin{center}  
  \caption{Data statistics of the four datasets used in all experiments.} 
  \renewcommand{\arraystretch}{1.1}
  \begin{tabular}{lcccc}
    \hline
    Dataset & Office  & Cell & Beauty & Sports \\
    \hline
    \#users & 4,905 & 27,879 & 22,363 & 35,598 \\
    \#products & 2,420 & 10,429 & 12,101 & 18,357 \\
    \#queries & 290 & 165 & 249 & 1,543 \\
    \#interactions & 53,258 & 194,439 & 198,502 & 296,337 \\
    \hline
  \end{tabular}  
  \label{tab:data_statistics}  
  \end{center} 
  \end{table}

\subsection{Existing PPS Models}
\label{sec:baseline_details}
HMPPS is aimed at reranking a limited number of candidate products filtered by existing PPS solutions. 
To verify the generalization of HMPPS, we choose six different, representative ID-based PPS approaches:
1) \textbf{HEM}~\cite{ai2017learning} jointly learns distributed embeddings for queries, products and users with a deep neural network;
2) \textbf{ZAM}~\cite{ai2019zero} utilizes an attention function over user purchased products to build user embeddings for product search;
3) \textbf{DREM}~\cite{ai2019explainable} constructs a knowledge graph to model the dynamic relationships between users and products in the latent space;
4) \textbf{DREM-HGN}~\cite{ai2021model} proposes to construct and train an intrinsic-explainable model for PPS with user-interaction data and knowledge graph;
5) \textbf{CAMI}~\cite{liu2022category} proposes a category-aware multi-interest model that learns multiple interest embeddings to encode diverse user preferences;
and 6) \textbf{UniSAR}~\cite{shi2024unisar} is trained jointly on personalized search and recommendation tasks to enhance the user preference comprehension for PPS.
We reproduce these methods based on their official codes to obtain the search results for HMPPS reranking.

To validate the superiority of HMPPS in content understanding, we compare it with multiple content-based approaches for PPS reranking. \textbf{RTM}~\cite{bi2021learning} utilizes a transformer to encode query, user and item reviews to obtain fine-grained interactions for PPS. 
\textbf{InstructRec}~\cite{zhang2023recommendation} leverages LLM to reason the relationship between textual user history, query and candidate item.
Since InstructRec aims to construct a general model for multiple recommendation and search tasks, it is trained on large amounts of synthesized data and evaluated underlying zero-shot setting. 
To keep fair comparison with HMPPS, which is trained with supervised setting, we reproduce the LLM-based PPS framework of InstructRec with a 3B language model and train the model on specific PPS datasets with its instruction tuning settings in its official report.

To validate the effectiveness of the two-stage training paradigm of HMPPS in user history selection, we also compare HMPPS with \textbf{QIN}~\cite{guo2023query}, which utilizes recommendation behaviors to augment search history and design a cascaded strategy to select user history. It is noteworthy that, to keep fair comparison with QIN reports, we apply \emph{leave-one-out} strategy to split data in the experiment.
 
\begin{figure}
  \centering 
  \includegraphics[width=0.48\textwidth]{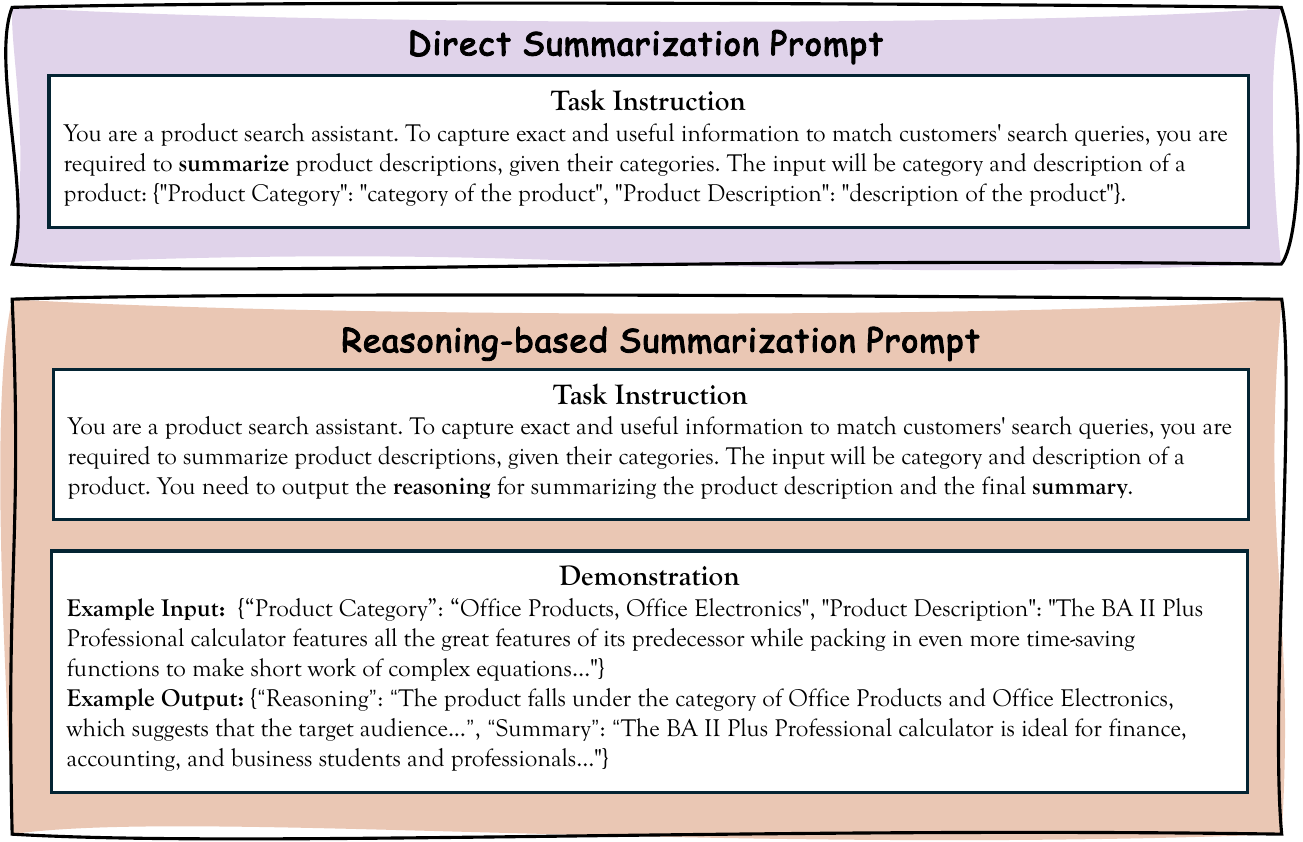}
  \caption{Example prompts for direct and reasoning-based description summarization.}
\label{fig:other_summarization_strategies}
\end{figure} 

\subsection{Implementation Details.}
\label{sec:implementation_details}
We take InternVL2-1B~\cite{chen2024internvl} as the MLLM backbone of HMPPS for most of our experiments.
We utilize Qwen2.5-14B~\cite{qwen2.5} to conduct description summarization and the number of core perspectives $K_d$ is set as 20 for all datasets.
The ID-based method $M_{ID}$, which generates the basic search results for HMPPS reranking, is UniSAR for all experiments, except for that in Table~\ref{tab:performance_comparison} which validates that HMPPS can adapt to any ID-based solutions for performance improvement.
The size of reranking candidate product set $K_p$ is set to 10. 
We sample 5 negative samples for training, where the number of simple negative samples $K_s^n$ is 2 and the number of hard negatives $K_h^n$ is 3. 
The product number $K_{s_1}$ of randomly selected user history for the first-stage of HMPPS is set as \{5, 7, 7, 7\} for Office, Cell, Beauty and Sports datasets, respectively.
The product number $K_{s_2}$ of user history selected based on the relevance to query and candidate product is set as \{2, 3, 3, 3\} for the second-stage of HMPPS.

Low-rank adaption (LoRA) is adopted for parameter-efficient finetuning of all components of MLLM, including the vision module, language module and the MLP layer. Benefitting from the remarkable generalization of MLLM, we train HMPPS on merely 10\% training data with only 1 epoch, batch size 1 and learning rate 0.0001 using AdamW optimizer on all datasets.

\subsection{Description Summarization Strategies}
\label{sec:other_description_summarization_strategies}
The specific prompts of the other two description summarization strategies in ablation study~\ref{description_summarization_ablation} are shown in Figure ~\ref{fig:other_summarization_strategies}. The direct summarization prompt instructs LLM to generate description summary directly without any demonstration and reasoning request. The reasoning-based summarization prompt requires LLM to reason before generating the final summary with one-shot demonstration, which can be seen as a simple version of perspective-guided summarization without an explicit declaration for specific perspectives.

\section{Other Related Work}
\noindent\textbf{User History Selection.}
Modeling user preference from their historical products helps improve product search and recommendation performance~\cite{pi2019practice}. 
However, redundancy and noise in user history cause accuracy decrease and latency burden of online systems. 
Therefore, various works focus on efficient user history selection. 
~\cite{pi2019practice, ren2019lifelong} leverage updatable, limited-size memory to store refined user history. 
~\cite{zhou2019deep, guo2019attentive} apply sequence modeling networks like GRU and LSTM to summarize user interest from user history. 
~\cite{zhou2018deep, ai2019zero, guo2023query, pi2020search} calculate attention to restrain the influence of query-irrelevant products.

HMPPS essentially also calculates attention to suppress redundancy and noise. Nevertheless, via the two-stage training paradigm, HMPPS obtains a migrated MLLM to extract robust multimodal representations for a more comprehensive matching between query and historical products, which is superior to those relying on ID-based embeddings and frozen semantic representations~\cite{ai2019zero, guo2023query}.
\end{document}